\documentclass[magisterska]{pracamgr}
\usepackage{graphicx}
\usepackage{amsmath,amssymb}
\usepackage[square,sort]{natbib}
\usepackage{lscape}
\usepackage{multirow}
\author{Damian Kwiatkowski}

\title{Progenitors of the Accretion-Induced Collapse}

\tytulang{Prekursory tzw. kolapsu indukowanego akrecj\k{a}}

\kierunek{Physics}

\zakres{Mathematical and computer modeling of physical processes}

\opiekun{Prof. dr hab. Krzysztof Belczy\'nski\\
Astronomical Observatory\\
  Warsaw University\\
\\
\textbf{Prof. dr hab. Marek Demia\'nski}\\
Insitute for Theoretical Physics\\
Faculty of Physics\\
Warsaw University\\
\\and
\textbf{dr Ashley J. Ruiter}\\
Research School of Astronomy and Astrophysics\\
Australian National University\\
Canberra, Australia}

\date{Wrzesie{\'n} 2015}

\dziedzina{ 
13.2 Physics
}


\begin{document}
	\maketitle
\begin{abstract}
%\\[2ex]
Naj\'swie\.zsze obliczenia dotycz\k{a}ce kolapsu indukowanego akrecj\k{a} dla tlenowo-neonowo- magnezowych bia{\l}ych kar{\l}\'ow \citep{b01}, prowadz\k{a}cego do powstania gwiazdy neutronowej, pokazuj\k{a}, \.ze zjawisko to mo\.ze wi\k{a}za\'c si\k{e} z emisj\k{a} obserwowalnego sygna{\l}u elektromagnetycznego. W zwi\k{a}zku z tym, w niniejszej pracy zaprezentowane s\k{a} teoretyczne cz\k{e}sto\'sci zaj\'scia tego zjawiska dla r\'o\.znych typ\'ow donor\'ow w uk{\l}adzie podw\'ojnym, tu\.z przed kolapsem. Wyniki uzale\.zniono r\'ownie\.z od typu, wieku oraz metaliczno\'sci syntetycznej galaktyki. Rezultaty prezentowane w tej pracy mog\k{a} wskaza\'c \'scie\.zk\k{e} dla przysz{\l}ych projekt\'ow obserwacyjnych, prowadz\k{a}c\k{a} do odnalezienia kandydat\'ow na uk{\l}ady, w kt\'orych zachodzi kolaps indukowany akrecj\k{a}. Moje prognozy bazuj\k{a} na symulacjach ewolucji uk{\l}ad\'ow podw\'ojnych, uwzgl\k{e}dniaj\k{a}cych najnowsze modele akrecji na bia{\l}e kar{\l}y. Wyniki wskazuj\k{a}, \.ze kolaps tlenowo-neonowo-magnezowych bia{\l}ych kar{\l}\'ow, akreuj\k{a}cych materi\k{e} z gwiazd w przerwie Hertzsprunga lub czerwonych olbrzym\'ow, jest najbardziej prawdopodobny. Typowe cz\k{e}sto\'sci zjawiska s\k{a} na poziomie $\sim 1-5\cdot 10^{-5} yr^{-1}$ w du\.zych galaktykach spiralnych oraz $\sim 10^{-4}-10^{-5} yr^{-1}$ w starych galaktykach eliptycznych. Prace przewiduj\k{a}, \.ze najsilniejszy sygna{\l} elektromagnetyczny ma pochodzi\'c z uk{\l}ad\'ow podw\'ojnych, w kt\'orych olbrzym pe{\l}ni funkcj\k{e} donora. Dla tego typu prekursor\'ow teoretyczna cz\k{e}stotliwo\'s\'c kolapsu wynios{\l}a $\sim 2.5\cdot 10^{-5} yr^{-1}$ w przypadku du\.zych galaktyk spiralnych oraz $\sim 1.3\cdot 10^{-5} yr^{-1}$ dla starych galaktyk eliptycznych.
\\[10ex]
\centerline{\textbf{Streszczenie (w j\k{e}zyku angielskim)}}
\\[2ex]
Recent calculations of accretion-induced collapse of an oxygen-neon-magnesium
white dwarf into a neutron star \citep{b01} allow for a potentially
detectable transient electromagnetic signal. Motivated by these results, I present
theoretical rates and physical properties of binary stars that can produce
accretion-induced collapse. The rates are presented for various types of host
galaxies (e.g. old ellipticals versus spirals) and are differentiated by the
donor star type (e.g. large giant star versus compact helium-rich donor).
Results presented in this thesis may help to guide near-future electromagnetic transient search
campaigns to find likely candidates for accretion-induced collapse events.
My predictions are based on binary evolution calculations that include the
most recent updates on mass accretion and secular mass growth of white dwarfs.
I find that the most likely systems that undergo accretion-induced collapse
consist of an ONeMg white dwarf with a Hertzsprung gap star or a red giant companion. Typical rates are of the
order $\sim 1-5\cdot 10^{-5} yr^{-1}$ and $\sim 10^{-4}-10^{-5} yr^{-1}$ in large spiral galaxies and old ellipticals, respectively.
The brightest electromagnetic signals are predicted for evolved giant type
donors for which I find rates of the order $\sim 2.5\cdot 10^{-5} yr^{-1}$ (large spirals) and $\sim 1.3\cdot 10^{-5} yr^{-1}$ (old
ellipticals).  
\end{abstract}
\textbf{Acknowledgements}\\ \\
Author of this work would like to acknowledge some certain scientists for numerous and very helpful remarks.\\ \\ First and most of all, I would like to offer my special thanks to Dr Ashley J. Ruiter from Mount Stromlo Observatory in Weston Creek, Australia, for help and guidance during the whole time I was working on the presented work.\\ \\ I would also like to express my very great appreciation to Mira Grudzi\'nska, PhD student from Astronomical Observatory of Warsaw University, for performing all needed \texttt{StarTrack} calculations.\\ \\
Finally, I would like to express my deep gratitude to Professor Krzysztof Belczy\'nski, my research supervisor, for his great patience, enthusiastic encouragement and useful critiques of this research work.
\tableofcontents

\chapter*{Introduction}
\addcontentsline{toc}{chapter}{Introduction}
Accretion-induced collapse is a process occuring in binary systems with white dwarfs (WDs) in which one of the WDs becomes a neutron star. Such events have never (knowingly) been observed, but theoretical preditions have existed for some time \citep{b68, b69, b70}. 

Before the collapse to a neutron star can be realized, neon and magnesium -- which are synthesized during oxygen (deflagration) burning -- undergo electron capture reactions which results in a decrease of the electron fraction. This in turn causes the electron-degenerate core to contract and heat up \citep{b67,b04}.
In the end, whether an AIC occurs or whether the core explodes thus hinges on a competition between energy generation in the core and the electron capture process, which depends on the central density of the WD \citep{b04}.

Thus, we are either left with a thermonuclear explosion (e.g. a type Ia supernova) or a core collapse event as the core approaches the Chandrasekhar mass limit. Type Ia supernovae are mostly expected to occur in carbon-oxygen WDs \citep{b02}, though there are some rare cases when AIC is predicted to originate from a CO WD that transformed to an ONeMg WD \citep{b05}. 

In this thesis, I am interested in the case of accretion-induced collapse events that occur in oxygen-neon magnesium WDs (ONeMg WD; \citet{b03,b04}). \\ \\
The remnant of an AIC event usually possesses too much angular momentum to collapse directly to nuclear density \citep{b06}.

Nevertheless, it may not be heated sufficiently to become convectively unstable, despite the high angular momentum.

Some AIC events will still produce hot and rapidly rotating NSs, thus it is generally accepted that AIC can explain some magnetars \citep{b07}. What is more, in such a case we should expect transient signals in X-ray and radio \citep{b99,b08}.\\ \\
Gamma-ray bursts, observed by BATSE, show an isotropic distribution on the sky, with a population decrease at low flux. This suggests, that they are at cosmological distances \citep{b12,b13,b14}. One of the accepted mechanisms of short GRBs could be the accretion-induced collapse, as well \citep{b15}. Particularly interesting within this scenario is a delayed magnetar formation through AIC or WD mergers \citep{b16,b17}.\\ \\
Other interesting objects that can be formed through the AIC channel are low-mass binary pulsars (LMBPs). Since the low-mass X-ray binary (LMXBs) population is 100 times smaller than LMBPs, AIC seems to be one potential reason for the huge discrepancy (\citealt{b09} and references therein). Also combinations of AIC and the tidal capture of NSs were suggested as an other scenario for the high density of LMBPs in globular clusters \citep{b10,b11}.\\ \\
Post-AIC evolution does not result in long-lived radio millisecond pulsars. Thus, it cannot explain most of the population of millisecond pulsars in globular clusters \citep{b18}. Nevertheless, it still gives a great contribution to formation of high B-field MSPs \citep{b19}.\\ \\
Despite the theoretical significance of AICs, no such event has ever been observed. One reason could be the fact that the optical transient is expected to be 5 magnitudes fainter compared with a typical type Ia supernova \citep{b01}. The energy of a typical AIC is predicted to be ~$10^{50}$ erg. The optical signature \citep{b55}
 
exhibits a peak absolute magnitude between ~-16 to -18, lasting for a few days. Apart from that, currently available theoretical models predict not more than one AIC per 100 exploding SNe Ia \citep{b20}.\\ \\
Fortunately, future observational surveys may finally detect events of such low rate and luminosity. One possibility could be Zwicky Transient Facility (ZTF; \citealt{b21}), All-Sky Automated Survey for Supernovae (ASAS-SN; \citealt{b22}), and the Large Synoptic Survey Telescope \citep{b23}. One must remember, though, that the strength of the optical and X-ray emission will strongly depend on the viewing angle. The strongest signal would come directly from the shocked region. But even with such strict constraints, \citet{b01} predict the detection of 13 AIC events per year with the ZTF aperture. \\ \\
When discussing progenitor scenarios, I should consider the two most fundamental cases. First one, which seems to be standard, is the case of a double-degenerate system, i.e. both stars in a binary are WDs. The degenerate donor will expand while losing mass, thus, the accretion rate is rapid on the dynamical timescale of the WD. Finally, it would be tidally disrupted and form a thick accretion disk around the more massive WD. If the accretion rate exceeds the Eddington limit by at least 10 \%, the accretor will never reach thermal equilibrium. Thus, a runaway nuclear burning will start near the star's surface. Such evolution occurs in a merger of two CO WDs with a combined mass larger than the Chandrasekhar mass, $M_{Ch} \approx 1.4 M_{\odot}$ and as a result a ONeMg WD would be formed. If the star continues to accrete, it may undergo AIC \citep{b04}.\\
Apart from this case, also close binary systems involving an ONeMg WD accreting from a companion via stable Roche-lobe overflow can lead to AIC. As mentioned previously, there is competition between the rapidity of energy generation from the deflagration burning, and the electron capture processes. In spite of this, for elements such as magnesium and neon, runaway fusion is much less likely than for carbon and oxygen (\citealt{b04} and references therein). 
Thus, formation of a neutron star is an expected 
outcome. Within this scenario, I can generally consider different donor types. A system where an ONeMg WD is accreting matter via stable Roche Lobe Overflow (RLOF) from a H-rich non-degenerate companion, non- or semi-degenerate helium-burning star, or a (degenerate) helium WD (e.g., \citealt{b100,b24,b25}). \\ \\
In this work, I will follow exemplary evolutions of each possible binary system with an ONeMg WD companion that subsequently undergoes the AIC process. Also, I will consider the evolution of stellar populations in elliptical (instantaneous burst of star formation at $t~=~0$) and spiral galaxies (continuous star formation until $t=10$ Gyr). It is also highly likely that rates of considered events would depend on metallicity \citep{b20,b28,b29,b30}. I present the results for three different metallicities: $Z_{\odot},~0.1Z_{\odot},~0.01Z_{\odot}$, assuming solar metallicity value of $Z=0.02$. Finally, I compare my results with previous studies.

\chapter{Model}\label{chap:chap1}

Evolutionary calculations presented in this work were performed using the \texttt{StarTrack} population synthesis code. The most comprehensive description of the input physics of the code can be found in \citeauthor{b31} \citeyearpar{b31,b32}, though several updates have since been made (e.g.: \citeauthor{b52} \citeyearpar{b52,b71}; \citealt{b36}).

Evolution of a single star is calculated starting from the ZAMS until a remnant is formed. Analytical formulae and evolutionary tracks used in the code are based on \citet{b33}. However, in the case of binary system evolution, physical processes can be quite complex and require additional treatment, e.g. mass transfer phases, common envelope (CE) evolution, SN kicks, magnetic braking, gravitational radiation or tidal interactions (\citealt{b32} for more details).
\section{Mass accretion}
Strong nova explosions impede the accumulation of H-rich matter onto a WD for very low mass transfer rates $<10^{-11}$ yr$^{-1}$ \citep{b34}. Above this threshold, an updated H-rich matter prescription that is based on \citet{b35} has been implemented \citet{b56}. Within this framework, fully efficient accumulation is only achieved within a narrow range of mass transfer rates. \\ \\
Sub-Chandrasekhar mass SN Ia explosions \citep{b66,b39,b72,b73} are possible when a He-rich shell is accumulated on a WD. Recent development in this field has opened a discussion about the details of He-rich matter accretion on the WD surface. New models proposed by \citeauthor{b37} \citeyearpar{b37,b38} have been introduced and used as a standard model in this thesis. In the most recent work, the authors show 4 regions on mass transfer rate- WD mass space, as well as fitted boundary formulae for them. Models presented in those publications demonstrate that the accumulation efficiency depends on mass of the WD and the accretion rate, especially in the \textit{strong flash regime}. This regime lies above the sub-Chandrasekhar mass SN~Ia region (called \textit{steady accumulation/double-detonation regime} in \citet{b37}. The most recent population synthesis study of sub-Chandrasekhar mass SNe~Ia taking into account such an accretion model is presented in \citet{b36}.

The donor stars in the above-described processes involve not only helium-burning stars; there is the possibility of helium WD (He WD) donors or hybrid WD donors, the latter which have a CO core and a He-rich mantle (e.g., \citealt{b39}, and references therein). Formation of a hybrid WD is possible when a red giant is deprived of its envelope in binary interactions. If the helium core does not commence He-burning reactions, a He WD is formed.\\ \\
If an ONeMg WD accretor does not pass in the parameter space which enables it to blow up as a sub-Chandrasekhar mass SN~Ia \citep{b65}, it steadily accretes mass until reaching the Chandrasekhar limit (\citealt{b40,b32}). Then it undergoes accretion-induced collapse to form a neutron star. 
 
\section{Orbital parameters}
Stellar populations, simulated in \texttt{StarTrack}, are randomly generated using initial mass function (IMF), which is a broken power law \citep{b42}. The ratio of primary and secondary mass in binary systems is chosen from a flat distribution in the range $q=0-1$ \citep{b43}. This is a canonical choice among population synthesis studies (e.g. \citealt{b44}). Initial orbital separation is computed from a distribution $\propto 1/a$ \citep{b45} and could be as high as $10^5~R_{\odot}$. It is also a widely accepted solution as it represents the local population of Hipparcos binaries \citep{b46}. Eccentricities are initially drawn from the distribution $\Psi(e)=2e$ \citep{b47}, which is a thermal-equillibrium distribution, in the range 0-1. The binary fraction is assumed to be 50\% which corresponds to 2/3 of all stars in the galaxy to be bound in binary systems. This assumption may be an overestimation for low-mass stars \citep{b48}, though is likely an underestimation for higher mass stars \citep{b74}.
\section{Common envelope} 
Binary evolution studies are still inconclusive about the mass-transferring binaries as well as the common envelope (CE) phase. The latter is actually one of the most problematic and least understood phenomena in theoretical astrophysics \citep{b75}.\\ 

The energy balance formula \citep{b49} is often used in binary population studies to estimate the orbital parameters following the CE phase:
\begin{equation}
\alpha_{CE}\left(\frac{GM_{d,f}M_{a}}{2a_{f}}-\frac{GM_{d,i}M_{a}}{2a_{i}}\right)=\frac{GM_{d,i}M_{d,env}}{\lambda R_{d,RL}}
\end{equation}
where $G$ is a gravitational constant, equals $6.674\cdot 10^{-11} m^{3}kg^{-1}s^{-2}$, $M_{d}$ and $M_{a}$ are masses of donor and accretor, respectively; $a$ is a binary separation; $R_{d,RL}$ - is Roche lobe radius of the donor; lower indices correspond to initial/final value ($_{i}/_{f}$); $_{env}$ - envelope of the accretor. The binding energy parameter of the envelope of the donor, $\lambda$, is referred to in this thesis as the "Nanjing" lambda (\citeauthor{b50} \citeyearpar{b50,b51}).

Detailed description of this parameter implementation into the current version of \texttt{StarTrack} is described in \citep{b52}. The $\alpha_{CE}$ parameter corresponds to the efficiency with which the orbital energy is transfered to the envelope of the mass-losing star. In this thesis I use the `standard model' for CE energy balance with 100\% efficiency, i.e. $\alpha$=1.
\section{Star Formation Rate} 
In order to study different types of galaxies, i.e. elliptical and spiral galaxies, I convolve resultant Delay Time Distribution (DTD) functions with relevant SFRs. It is common practice in elliptical galaxy studies to use an instantaneous burst SFR at $t=0$. For my model spiral galaxy, I adopt a SFR that is constant until $t=10$ Gyr. After that, no stars are formed. My study involves populations evolved up to 13.7 Gyr, which corresponds to the approximate age of the Universe. The mass of stars formed in both model galaxies is equal to $6\cdot 10^{10}~M_{\odot}$. This is fully consistent with the stellar mass of the Milky Way (MW; \citealt{b53}).

\chapter{Results}\label{chap:chap2}
\section{Evolution of AIC progenitor binaries} 
In this section, I present typical evolutionary scenarios that lead to accretion-induced collapse. Each subsection is named after the secondary star in the system at the onset of AIC event. Metallicity for the MS star with $M\leq 0.7 M_{\odot}$ is from the 1\% solar metallicity model only, since no such system occured in the other $Z$ models. For all subsequent evolutionary paths presented in this section, I only present solar metallicity models, i.e. $Z=0.02$, because systems with the same initial parameters, but different metallicity pass through the next evolutionary phases within the same timeline.

Mass transfer rates statistics is also presented in Table 2.1. I assume that during AIC the kick velocity is equal to zero. Evolution of primary $M_1$ and secondary $M_2$ mass, as well as accretion rate $\dot{M}$ and semi-major axis with eccentricity ($a,e$) is shown in Figures 2.1-2.11. Figures 2.12-2.13 depict the evolution of the AIC progenitors that are predicted to produce a significant electromagnetic transient signal, i.e. binaries with RG and AGB donors.
\subsection{Main Sequence star with $M\leq 0.7 M_{\odot}$}
\begin{figure}
\centering
\includegraphics[width=1.0\textwidth]{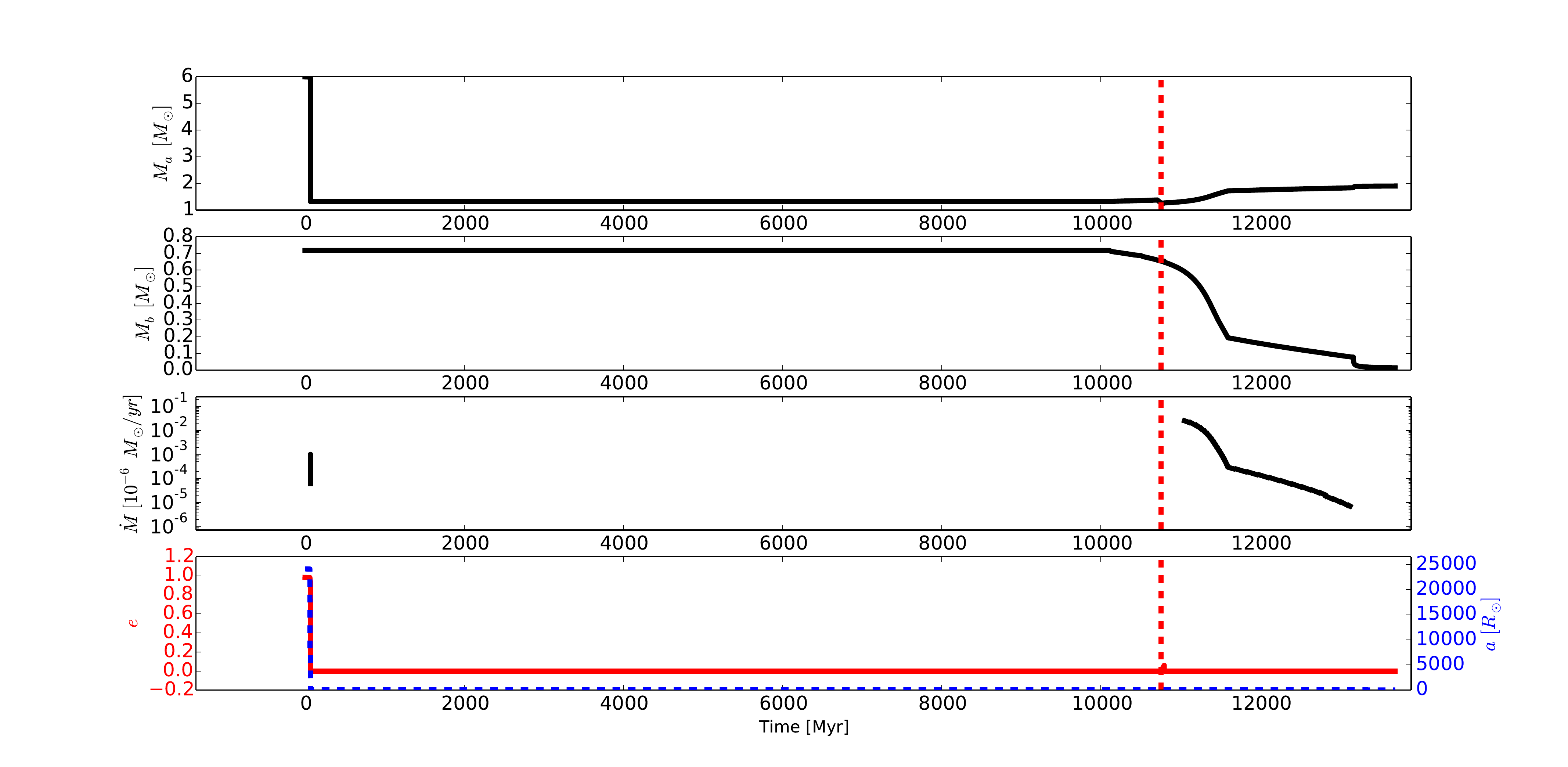}
\caption{Evolution of accretor mass $M_1(t)$, donor mass $M_2(t)$, mass transfer rate $\dot{M}(t)$, semi-major axis $a$ (blue dashed line) and eccentricity $e$ (red line) of an orbit for MS donor with $M\leq 0.7 M_{\odot}$. A red dashed vertical line represents a moment when AIC occurs.}
\label{fig1}
\end{figure}
Binary evolution starts with two components on the Zero Age Main Sequence (ZAMS) with $M_1=5.98M_{\odot}$, $R_1=3.95R_{\odot}$ (primary) and $M_2=0.71M_{\odot}$, $R_2=0.64R_{\odot}$ (secondary). Semi-major axis of the initial orbit is $a=2.4\cdot 10^4 R_{\odot}$ and its eccentricity equals to $e=0.98$ (with the orbital period $P=168000$ days). After 61 Myr the primary finishes core hydrogen burning, not changing its mass significantly. Then it enters the Hertzsprung gap. At the same time, eccentricity of the orbit starts to drop, as the radius of primary star increases and because of tidal interactions. Only within about 0.3 Myr, the primary enters the first giant branch only for 10,000 years and becomes a core helium-burning star. This phase is also relatively short and after 7 Myr primary becomes an early AGB star of a radius $R_1=134.6R_{\odot}$. The orbit becomes fully circularized ($a=741.5R_{\odot}$. At $t=68.92$ Myr, the primary starts to pulse thermally, being as large as $R_1=295R_{\odot}$. After about 20 years the primary passes through a CE phase and ejects its envelope, becoming an ONeMg WD of $M_1=1.32M_{\odot}$ mass and $R_1=2000$ km. The semi-major axis drops down to $a=5.29R_{\odot}$ (orbital period: $P=1$ day). Now the secondary evolves along its MS and finally at $t=10.1$ Gyr the binary system initiates mass transfer on a nuclear timescale as the secondary ($M_b=0.71 M_{\odot}$) overfills its Roche lobe and the orbital radius drops to $a=2.1R_{\odot}$. Until $t=10.4$ Gyr, the secondary loses mass and becomes deeply convective. Accretion-induced collapse occurs at $t=10.7$ Gyr. Before this event, the primary mass was $M_1=1.38 M_{\odot}$ and lost $0.11 M_{\odot}$ during neutronization. Its radius dropped from $R_1=2000$ to $9.7$ km.\\ \\ 
Evolution of this particular system was traced further on. The orbital period of the binary decreases to $P=70$ min. Nuclear MT terminates at $t=11.4 Gyr$ and after 1.3 Gyr, 0.43 $M_{\odot}$ was accreted from the donor. At $t=11.5$ Gyr, a thermal RLOF starts. After 1.5 billion years, the secondary is fully deprived of its envelope and becomes a helium WD of $M_2=0.1M_{\odot}$. The orbit increases during next 500 Myr up to $a=1.01R_{\odot}$ and then again its radius drops down to $0.5 R_{\odot}$ at $t=13.1$ Gyr as the mass transfer stops. Evolution ends at a Hubble time, when the binary system parameters are as follows: $M_1=1.84M_{\odot}$, $M_2=0.08M_{\odot}$, $a=0.2 R_{\odot}$ and the orbital period is $P=9.5$ min.
\subsection{Main Sequence star with $M\geq 0.7 M_{\odot}$}
\begin{figure}
\centering
\includegraphics[width=1.0\textwidth]{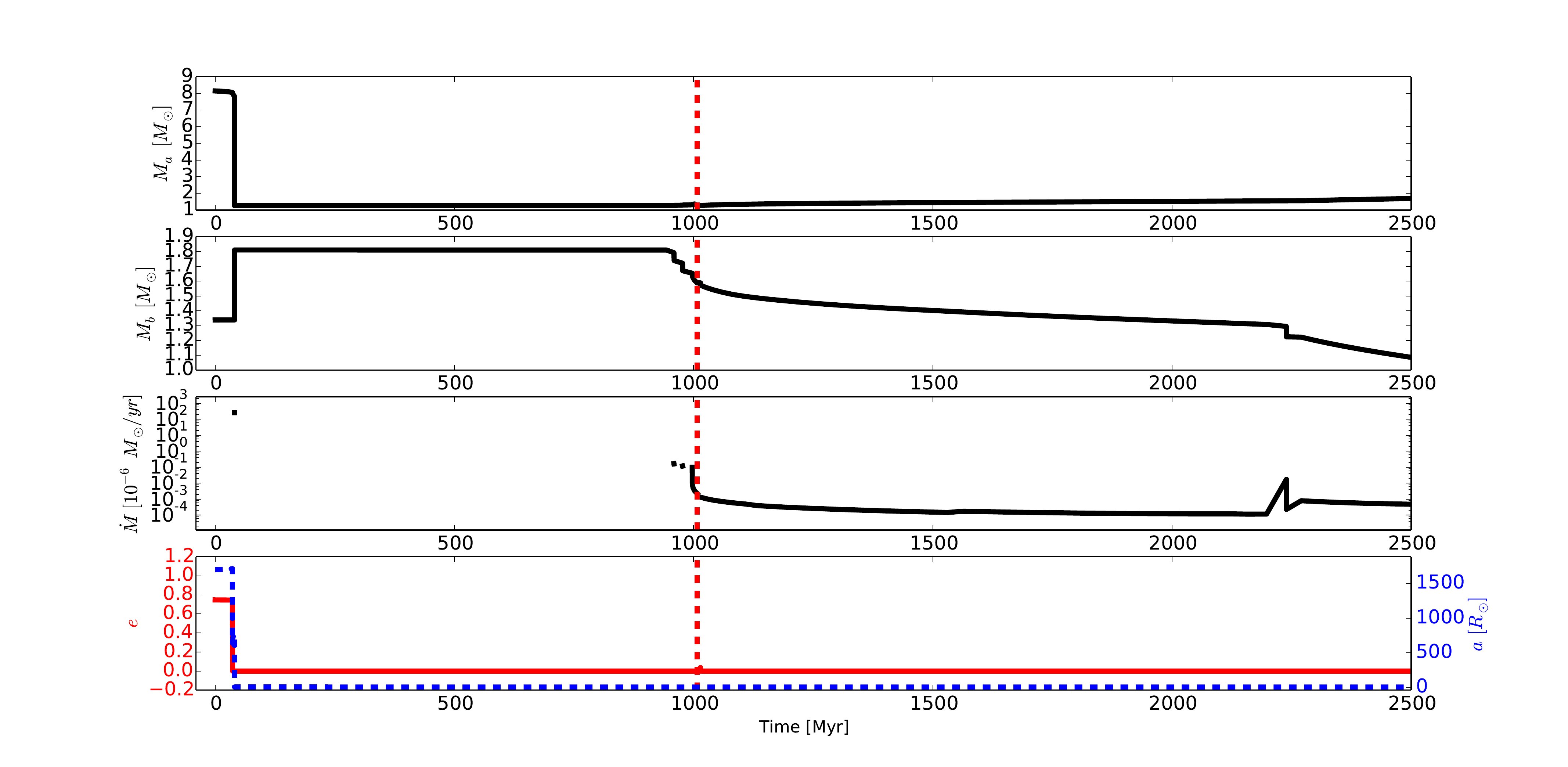}
\caption{Evolution of accretor mass $M_1(t)$, donor mass $M_2(t)$, mass transfer rate $\dot{M}(t)$, semi-major axis $a$ (blue dashed line) and eccentricity $e$ (red line) of an orbit for MS donor with $M\geq 0.7 M_{\odot}$. A red dashed vertical line represents a moment when AIC occurs.}
\end{figure}
Evolution starts with ZAMS stars of $M_1=8.15M_{\odot}$ (primary), $M_2=1.34M_{\odot}$ (secondary). The initial semi-major axis is $a=1.7 10^3 R_{\odot}$ and its eccentricity is $e=0.75$ (orbital period is $P=2635$ days). At $t=36.1$ Myr the primary becomes an HG star and the primary expands strongly. Tidal forces start to circularize the orbit. After 0.1 Myr, primary passes through giant branch, being as large as $R_1=275.5 R_{\odot}$. Within 20,000 years after that, the orbit becomes fully circular of $a=672.2 R_{\odot}$. 0.3 Myr later the primary starts to burn helium inside the core and subsequently fills its Roche lobe. The orbital radius drops down to $a=4.84 R_{\odot}$ and the primary becomes an HG naked He star as the binary passes through CE phase. After 0.02 Myr, the primary fills its Roche lobe and mass transfer on a thermal timescale begins. Within the next 0.02 Myr, the primary enters the giant branch for a few hundred years. Then, at $t=40.67 Myr$ it becomes an ONeMg WD, decreasing its stellar radius from $R_1=106.7 R_{\odot}$ to $R_1= 2782$ km . A thermal MT is ongoing until $t=977.1 Myr$ when it becomes nuclear. At $t=997.4$ Myr secondary fills its Roche lobe, while still being an MS star with $M_2=1.74 M_{\odot}$. Finally, at $t=1.0$ Gyr an AIC event occurs. The primary becomes a neutron star of mass $M_1=1.26 M_{\odot}$, losing $0.1 M_{\odot}$ through the process. The radius of the star $R_1=9.74$ km. Until this point, the secondary lost 0.3 $M_{\odot}$ via mass transfer. \\ \\
After that, RLOF still occurs on a thermal timescale and the secondary becomes an HG star at $t=2.24$ Gyr. When the mass transfer finally stops, the orbital radius is $a=5.21 R_{\odot}$, which corresponds to a period $P=19.9$ hours. The stellar masses are $M_1=1.55 M_{\odot}$, and $M_2=1.23 M_{\odot}$.
\subsection{Hertzsprung gap star}
\begin{figure}
\centering
\includegraphics[width=1.0\textwidth]{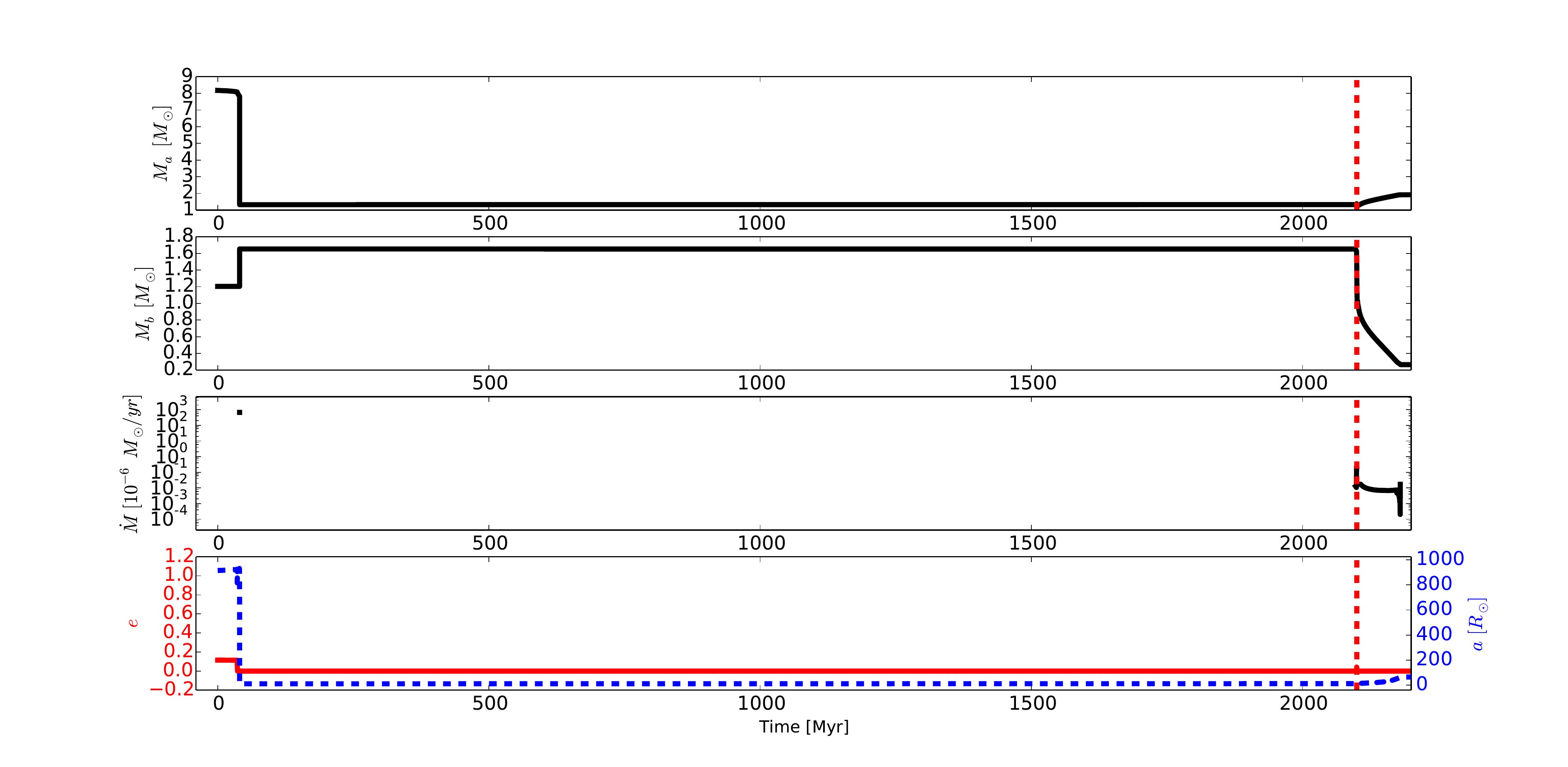}
\caption{Evolution of accretor mass $M_1(t)$, donor mass $M_2(t)$, mass transfer rate $\dot{M}(t)$, semi-major axis $a$ (blue dashed line) and eccentricity $e$ (red line) of an orbit for HG donor. A red dashed vertical line represents a moment when AIC occurs.}
\end{figure}
Binary evolution starts with ZAMS stars of the following masses and radii: $M_1=8.2 M_{\odot}$, $R_1=8.26 R_{\odot}$, $M_2=1.2 M_{\odot}$, $R_2=1.15 R_{\odot}$. The initial orbital parameters are as follows: $a=915.2 R_{\odot}$, $e=0.1$ (with orbital period $P=1048$ days). After 35.8 Myr, the primary becomes an HG star. Tidal forces circularize the orbit, as the primary significantly expands. Within 0.2 Myr, star becomes a first giant branch star, but only for 0.25 Myr. At $t=36.0$ Myr primary enters CHeB and shortly after that the orbit becomes fully circular with semi-major axis $a=817 R_{\odot}$ which corresponds to $P=889$ days. After 4.3 Myr the primary becomes an early AGB star, with the orbital radius still increasing. At $t=40.4$ Myr the binary passes through a CE phase and the naked He star returns to the Hertzsprung gap. Its mass drops from $M_1=7.8 M_{\odot}$ to $M_1=2.2 M_{\odot}$. At this time, the orbital radius is $a=10.3 R_{\odot}$. Shortly after, the primary fills its Roche lobe and transfers mass on a thermal timescale. Within a hundred years the primary passes through the giant branch, slightly decreasing the orbital radius to $a=9.73 R_{\odot}$ and finally becomes an ONeMg WD. Its mass equals to $M_1=1.32 M_{\odot}$ and stellar radius decreased from $R_1=87.7 R_{\odot}$ to $R_1=1947$ km. Mass transfer stops at this time. At $t=2.09$ Gyr, the secondary becomes an HG star and 30 Myr after that, fills its Roche lobe. This starts another mass transfer phase, but on a nuclear timescale. Within the next 0.1 Myr, an accretion-induced collapse occurs and the primary's mass changes from $M_1=1.38 M_{\odot}$ to $M_1=1.26 M_{\odot}$, while transforming into a neutron star. \\ \\
As soon as the mass transfer stops, i.e. at $t=2.1$ Gyr, the secondary star enters the giant branch and a thermal timescale RLOF occurs, lasting for 80 Myr. During this period, the primary accreted $0.7 M_{\odot}$ from its stellar companion and the orbital radius increased to $a=65 R_{\odot}$. Final parameters of the binary stars are as follows: $M_1=1.91 M_{\odot}$, $M_2=0.26 M_{\odot}$, $R_1=9.74$ km, $R_2=0.018 R_{\odot}$.
\subsection{I Giant Branch star}
\begin{figure}
\centering
\includegraphics[width=1.0\textwidth]{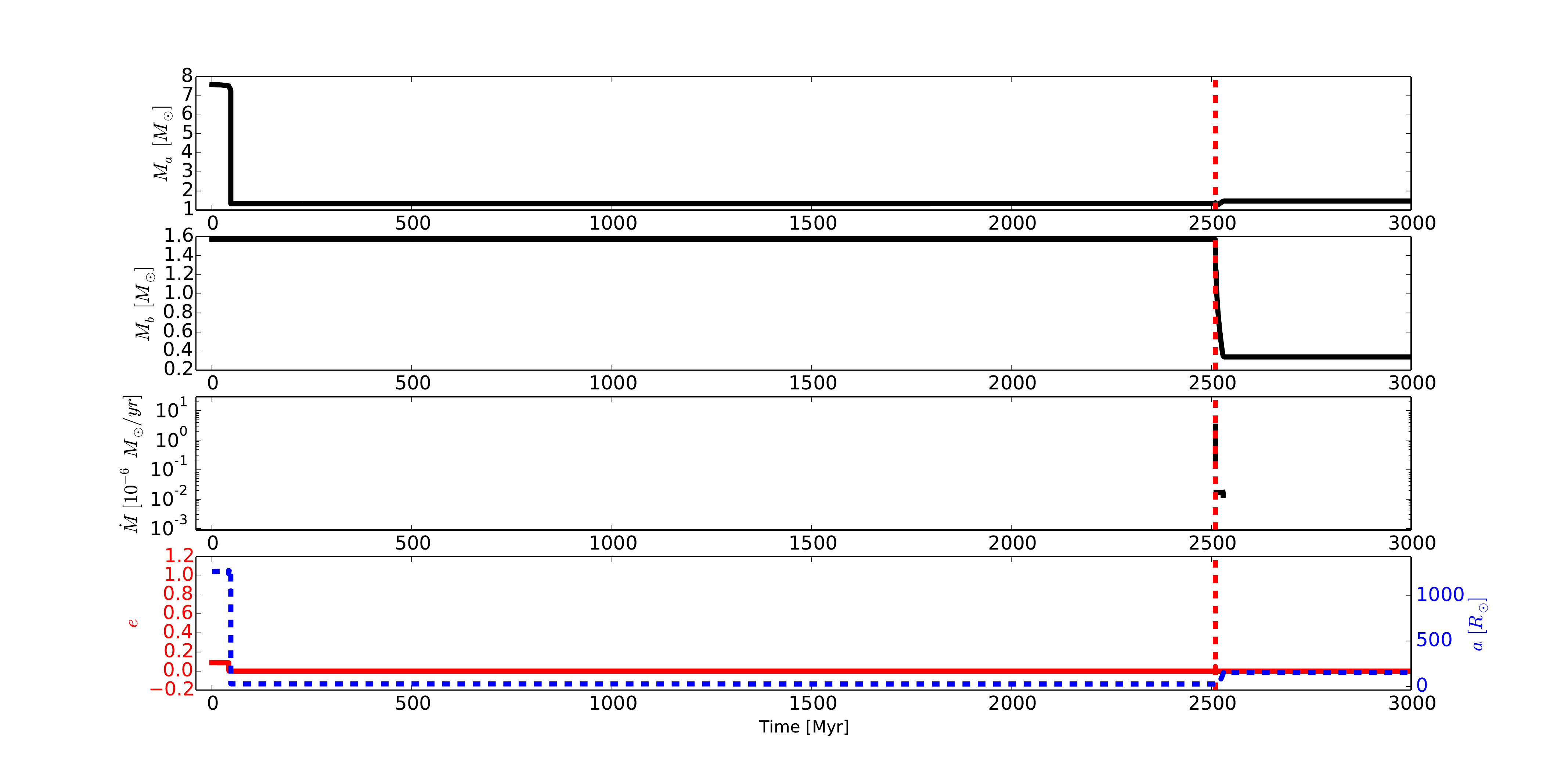}
\caption{Evolution of accretor mass $M_1(t)$, donor mass $M_2(t)$, mass transfer rate $\dot{M}(t)$, semi-major axis $a$ (blue dashed line) and eccentricity $e$ (red line) of an orbit for Red Giant donor. A red dashed vertical line represents a moment when AIC occurs.}
\end{figure}

Binary companions at ZAMS have the following masses and radii: $M_1=7.6 M_{\odot}$, $R_1=7.9 R_{\odot}$, $M_2=1.6 M_{\odot}$, $R_2=1.5 R_{\odot}$. The initial semi-major axis is $a=1266 R_{\odot}$ and its eccentricity is $e=0.09$ (with orbital period $P=1726$ days). After 41.61 Myr, the primary star becomes an HG star and its radius increases by a factor of 6. At $t=41.74$ Myr, the star enters the giant branch and expands up to $R=112.2 R_{\odot}$. Also the orbit begins to circularize, due to tidal forces and expansion of the primary. After 37,000 years the primary starts to burn helium inside the core and the orbit finally becomes circular. At $t=46.9$ Myr the star climbs the AGB and thermal pulsations start within 0.2 Myr. Then primary then starts to overfill its Roche lobe. Now its radius has grown to $R_1=538 R_{\odot}$. This evolutionary phase is very short, and after a CE phase the star ejects the envelope and becomes an ONeMg WD of radius $R_1=2.44\cdot 10^{-3} R_{\odot}$ and mass $M_1=1.33 M_{\odot}$. The orbital radius drops to $a=29.3 R_{\odot}$, which corresponds to $P=10.8$ days. At $t=2.36$ Gyr, the secondary enters the Hertzsprung gap. Within the next 52 Myr, the star becomes a giant . At $t=2.51$ Gyr, a rapid mass transfer via RLOF begins. Finally, after 0.3 Myr, an AIC occurs. During this process, the primary becomes a neutron star of $M_1=1.26 M_{\odot}$ and $R_1=9.74$ km.\\ \\
After that, until $t=2.53$ Gyr, the primary accretes matter from the donor. Subsequently, the secondary becomes a small He WD. Final parameters of this system are following: $M_1=1.47 M_{\odot}$, $ M_2=0.34 M_{\odot}$, $R_1=9.74$ km, $R_2=0.017 R_{\odot}$, $a=157.5 R_{\odot}$, $P=170.3$ days.
\subsection{Early AGB star}
\begin{figure}
\centering
\includegraphics[width=1.0\textwidth]{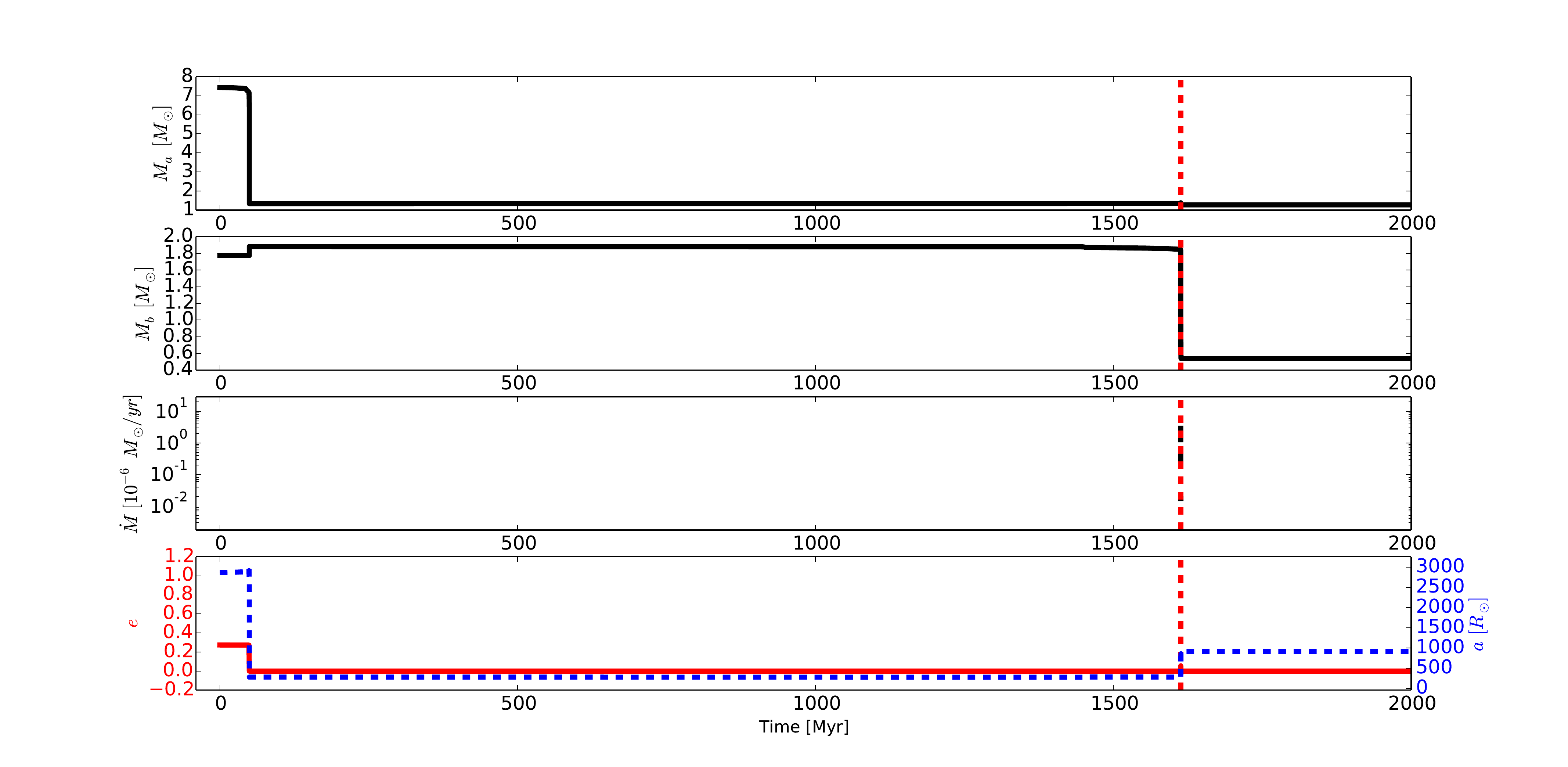}
\caption{Evolution of accretor mass $M_1(t)$, donor mass $M_2(t)$, mass transfer rate $\dot{M}(t)$, semi-major axis $a$ (blue dashed line) and eccentricity $e$ (red line) of an orbit for early AGB donor. A red dashed vertical line represents a moment when AIC occurs.}
\end{figure}

A typical progenitor of this composition starts on the ZAMS with the following stellar and orbital parameters: $M_1=7.43 M_{\odot}$, $R_1=7.8 R_{\odot}$, $M_2=1.77 M_{\odot}$, $R_2=1.55 R_{\odot}$, $a=2871 R_{\odot}$, $e=0.27$ (orbital period $P=5877$ days). At $t= 43.4$ Myr, the primary enters the Hertzsprung gap and after next 0.14 Myr becomes a giant. This phase is also very short and at $t=48.9$ Myr, the star becomes a CHeB star of $R_1=206.9 R_{\odot}$ being only slightly less massive than at ZAMS. Within the next 0.5 Myr, the primary ascends the AGB  and the orbit circularizes due to tidal interactions. Then it becomes an ONeMg WD through a CE phase. The orbital radius is now $a=282.3 R_{\odot}$. The stellar components have the following masses and radii: $M_1=1.34 M_{\odot}$, $R_1=2.4\cdot 10^{-3} R_{\odot}$, $M_2=1.88 M_{\odot}$, $R_2=1.6 R_{\odot}$. During $t=1.4-1.6$ Gyr, the secondary climbs the HG, giant branch, CHeB and finally becomes an early AGB star. In spite of this fact, no significant changes have been observed in the evolutionary data, apart from secondary growth from $R_2=3.4 R_{\odot}$ to $R_2=23.1 R_{\odot}$. And then mass transfer starts on a thermal timescale. As a result, the accretor becomes a neutron star of $M_1=1.26 M_{\odot}$, losing $0.11 M_{\odot}$ during electron capture. The orbital radius is now $a=406.1 R_{\odot}$.\\ \\
Within the century, the binary continues to transfer matter on a nuclear timescale. The secondary shortly begins to experience thermal pulses, but finally is deprived of its envelope and becomes a CO WD. At this point, the binary system can be characterised by this set of parameters: $M_1=1.27 M_{\odot}$, $R_1=9.47$ km, $M_2=0.54 M_{\odot}$, $R_2=0.013 R_{\odot}$.
\subsection{Thermally pulsing AGB star}
\begin{figure}
\centering
\includegraphics[width=0.7\textwidth]{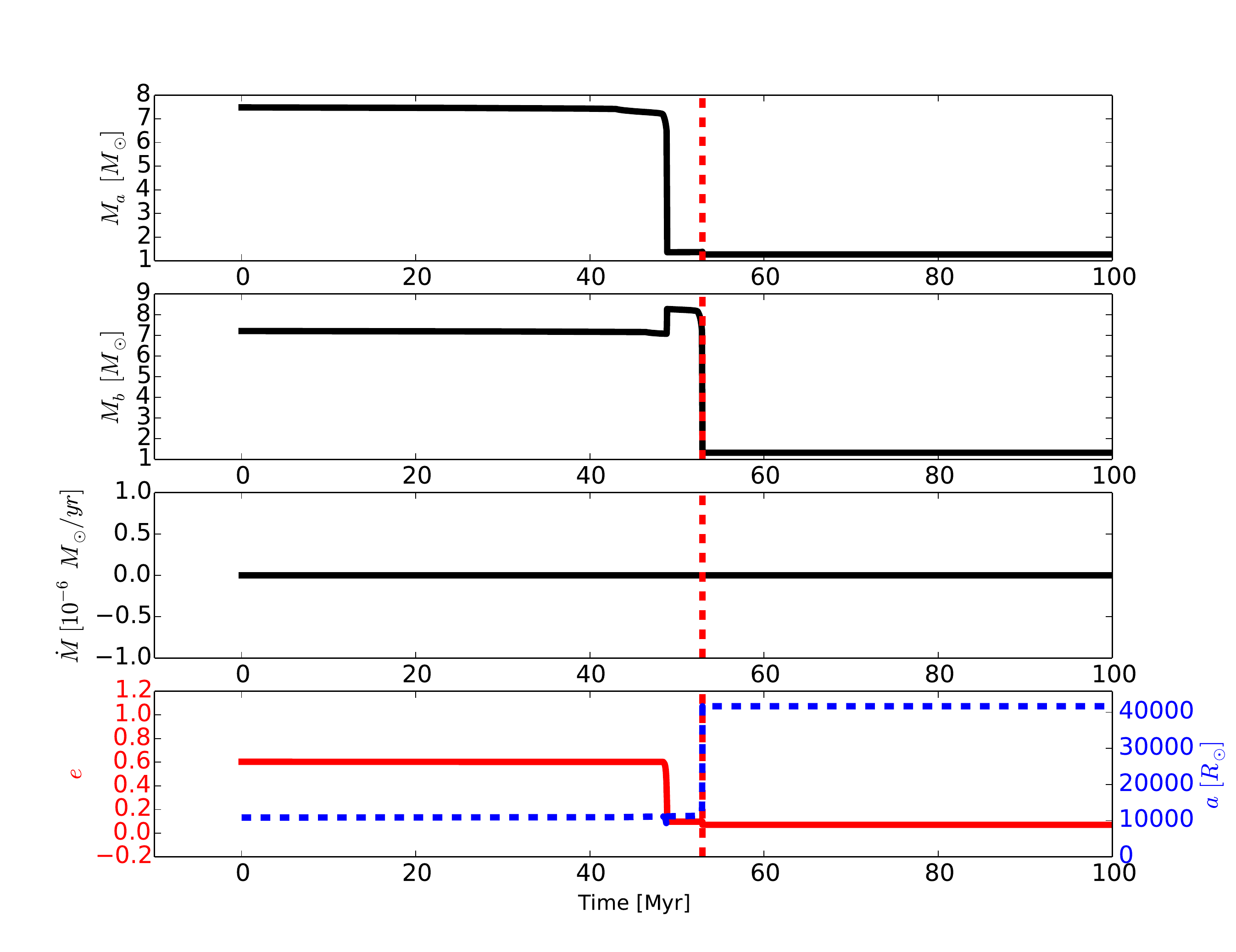}
\caption{Evolution of accretor mass $M_1(t)$, donor mass $M_2(t)$, mass transfer rate $\dot{M}(t)$, semi-major axis $a$ (blue dashed line) and eccentricity $e$ (red line) of an orbit for thermally pulsing AGB star. A red dashed vertical line represents a moment when AIC occurs.}
\end{figure}
Binary evolution starts with ZAMS stars of the following masses and radii: $M_1=7.48 M_{\odot}$, $R_1=7.83 R_{\odot}$, $M_2=7.21 M_{\odot}$, $R_2=7.41 R_{\odot}$. The semi-major axis of initial orbit is $a=1.08\cdot 10^4 R_{\odot}$ and its eccentricity $e=0.6$ (orbital period $P=93.5$ years). After 42.7 Myr, the primary passes through the Hertzsprung gap, enters the giant branch and starts helium burning inside the core. Between $t=46.3-46.4$ Myr, the secondary passes the same evolutionary path as the primary until formation of a CHeB star. At this point, the stellar and orbital parameters are as follows: $M_1=7.29 M_{\odot}$, $R_1=56.62 R_{\odot}$, $M_2=7.15 M_{\odot}$, $R_2=224.35 R_{\odot}$, $a=11028.6 R_{\odot}$, $e=0.6$, $P = 96.7$ years. Then at $t=48.2-48.4$ Myr, the primary becomes an early AGB star and starts thermal pulsations. After 0.4 Myr, it becomes an ONeMg WD. Later, at $t=52.1-52.4$ Myr, the secondary passes through the AGB and starts thermal pulses.  After the next 0.5 Myr, the primary becomes a neutron star. Before this event, its mass was $M_1=1.38 M_{\odot}$ and the radius was equal to $R_1=1.24\cdot 10^{-4} R_{\odot}$. After becoming a neutron star, the primary lost $0.11 M_{\odot}$ and its radius decreased to $R_1=1.4 \cdot 10^{-5} R_{\odot}$. The semi-major axis is 3 times larger and the eccentricity - 6 times smaller. 8,000 years after the AIC event, the secondary changed into an ONeMg WD. Final set of stellar and orbital parameters is the following: $M_1=1.26 M_{\odot}$, $R_1=9.74$ km, $M_2=1.33 M_{\odot}$, $R_2=1947$ km, $a=4.17\cdot 10^4 R_{\odot}$, $e=0.07$, $P=1676$ years.
\subsection{MS naked He star}
\begin{figure}
\centering
\includegraphics[width=1.0\textwidth]{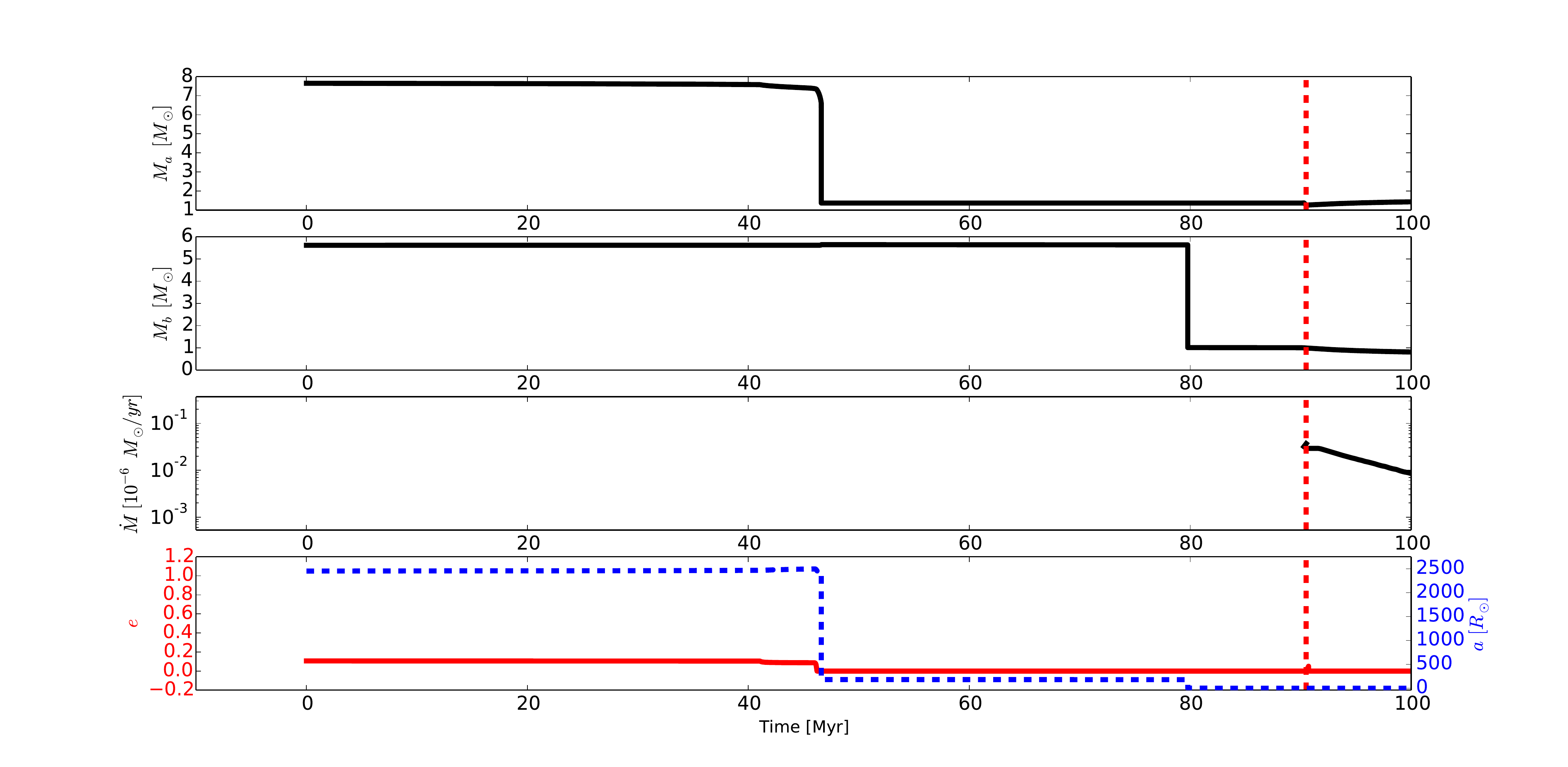}
\caption{Evolution of accretor mass $M_1(t)$, donor mass $M_2(t)$, mass transfer rate $\dot{M}(t)$, semi-major axis $a$ (blue dashed line) and eccentricity $e$ (red line) of an orbit for naked MS He donor. A red dashed vertical line represents a moment when AIC occurs.}
\end{figure}
Evolution starts with ZAMS primary of mass $M_1=7.65 M_{\odot}$, radius $R_1=7.94 R_{\odot}$ and secondary of mass $M_2=5.61 M_{\odot}$, radius $R_2=7.87 R_{\odot}$. The orbital parameters are the following: $a=2452.78 R_{\odot}$, $e=0.1$ (orbital period: $P=3865.8$ days). The primary becomes an HG star at $t=40.9$ Myr, enters the giant branch 10,000 years later and its radius grows up to $R_1=114.2 R_{\odot}$. At this point, the orbit starts to circularize, due to tidal forces. At $t=41.0$ Myr, the primary starts to burn helium inside the core. Within 5 Myr, it becomes an early AGB star and after the next 0.2 Myr the star starts thermal pulsations. The primary radius is as high as $R_1=516.1 R_{\odot}$ and the orbit becomes fully circularized. At $t=46.6$ Myr, the star becomes an ONeMg WD of mass $M_1=1.37 M_{\odot}$ and radius $R_1=1.3\cdot 10^{-3} R_{\odot}$. During the period of $t=79.43-79.78$ Myr, the secondary passes through MS, HG and giant branch (as described above). At $t=90.3$ Myr, the secondary fills its Roche lobe and nuclear timescale mass transfer starts. Then, the primary becomes a neutron star through AIC. Right before collapse, binary parameters were as follows: $M_1=1.37 M_{\odot}$, $R_1=5939$ km, $M_2=1.0 M_{\odot}$, $R_2=0.25 R_{\odot}$, $a=0.7 R_{\odot}$. After this event, only mass and radius of the primary changed significantly to: $M_1=1.26 M_{\odot}$, $R_1=9.74$ km.\\ \\
15 Myr later, the secondary enters the HG and overfills its Roche lobe. Accretion onto the primary occurs for 3.5 Myr. After that, the secondary star becomes a CO WD and final parameters are: $M_1=1.56 M_{\odot}$, $R_1=9.74$ km, $M_2=0.6 M_{\odot}$, $R_2=0.012 R_{\odot}$, $a=0.64$, $P=58.54$ min.
\subsection{HG naked He star}
\begin{figure}
\centering
\includegraphics[width=1.0\textwidth]{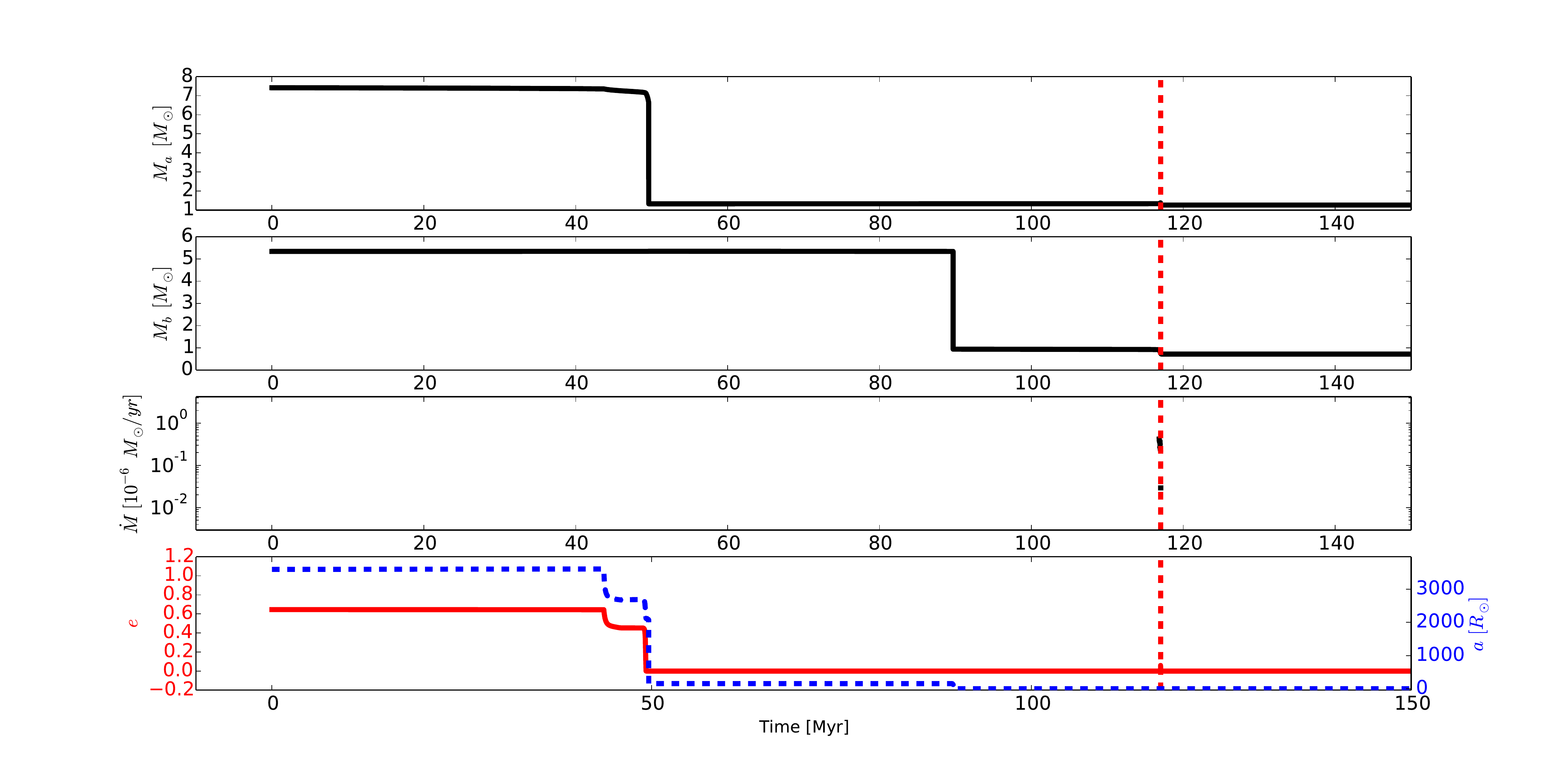}
\caption{Evolution of accretor mass $M_1(t)$, donor mass $M_2(t)$, mass transfer rate $\dot{M}(t)$, semi-major axis $a$ (blue dashed line) and eccentricity $e$ (red line) of an orbit for naked HG He donor. A red dashed vertical line represents a moment when AIC occurs.}
\end{figure}
This system evolution starts from ZAMS with primary parameters of $M_1=7.42 M_{\odot}$, $R_1=7.8 R_{\odot}$ and secondary parameters $M_2=5.34 M_{\odot}$, $R_2=7.80 R_{\odot}$. The semi-major axis is $a=3613.0 R_{\odot}$ and eccentricity $e=0.65$ (orbital period: $P=$7000 days). At $t=43.5-49.0$ Myr, the primary passes through MS, HG and AGB. Circularization of the orbit (caused by tidal interactions) also occurs as soon as the primary becomes an early AGB star. But within 0.01 Myr also thermal pulsations of this AGB star begin. At $t=49.6$ Myr, star overfills its Roche lobe. Right after that, an ONeMg WD is formed during CE event. Radius of the orbit decreases to $a=157.4 R_{\odot}$. 40 Myr later, the secondary enters the HG gap and giant branch, subsequently returning on the MS as a naked He star. The orbit contracts from $a=129.4 R_{\odot}$ to $a=1.31 R_{\odot}$. Also at this point, stable RLOF starts to occur. At $t=114.4$ Myr, the secondary enters the Hertzsprung gap. 2.6 Myr later, the primary undergoes neutronization through the AIC channel. Before that: $M_1=1.38 M_{\odot}$, $R_1=1468$ km, $M_2=0.77 M_{\odot}$, $R_2=0.5 R_{\odot}$, $a=1.51 R_{\odot}$. After neutronization, only the primary parameters changed to: $M_1=1.26 M_{\odot}$, $R_1=9.74$ km.  At $t=117.1$ Myr, the secondary is depleted from envelope mainly through accretion and becomes a CO WD. Final parameters of this binary system are: $M_1=1.26 M_{\odot}$, $R_1=9,74$ km, $M_2=0.72 M_{\odot}$, $R_2=0.011 R_{\odot}$, $a=1.67 R_{\odot}$. 
\subsection{Giant Branch naked He star}
\begin{figure}
\centering
\includegraphics[width=1.0\textwidth]{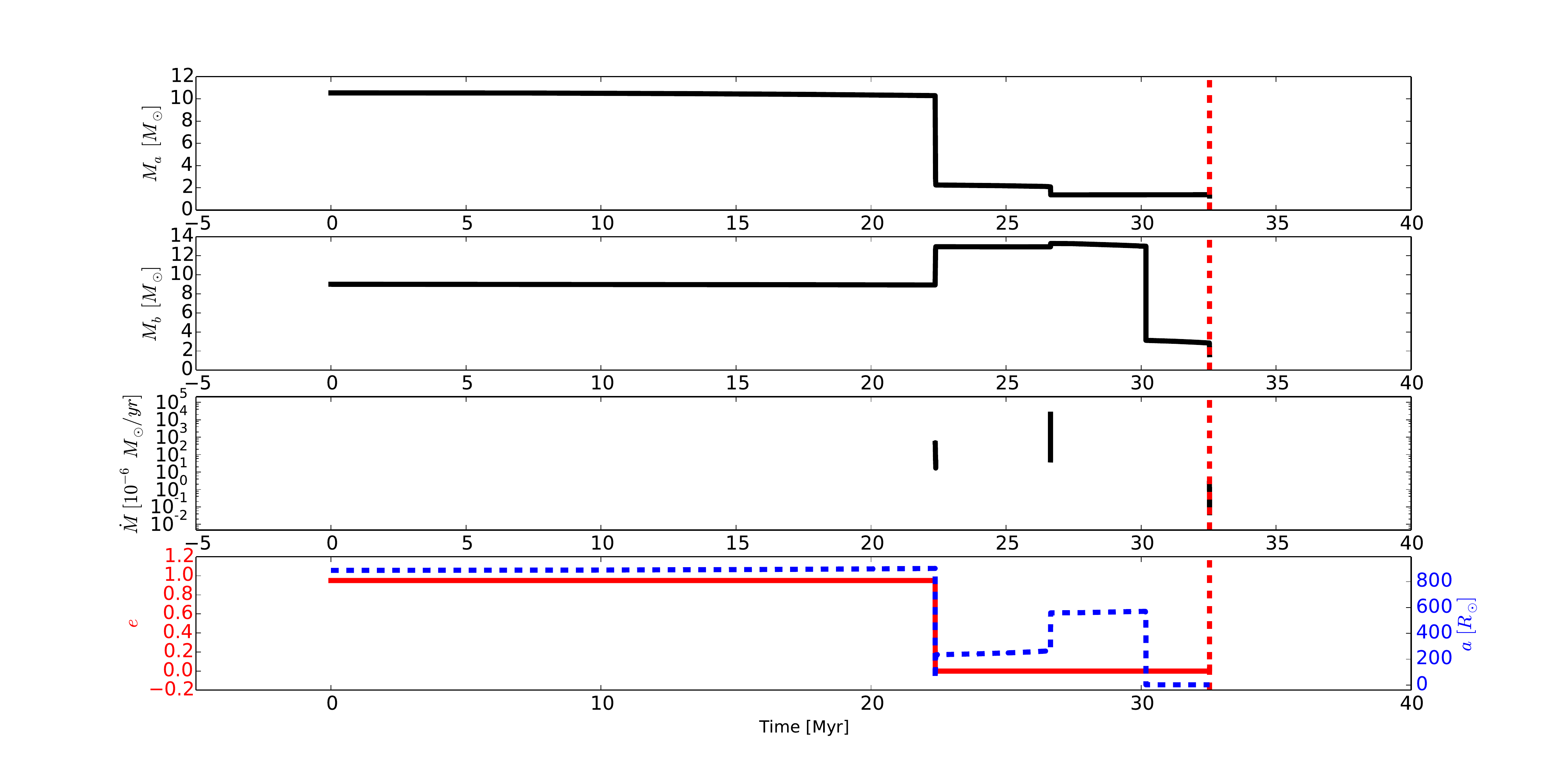}
\caption{Evolution of accretor mass $M_1(t)$, donor mass $M_2(t)$, mass transfer rate $\dot{M}(t)$, semi-major axis $a$ (blue dashed line) and eccentricity $e$ (red line) of an orbit for Giant Branch naked He donor. A red dashed vertical line represents a moment when AIC occurs.}
\end{figure}
Progenitor binary at ZAMS starts evolution with the following parameters: $M_1=10.5 M_{\odot}$, $R_1=9.7 R_{\odot}$, $M_2=9.0 M_{\odot}$, $R_2=6.2 R_{\odot}$. Initial semi-major axis is $a=887,9 R_{\odot}$ and eccentricity $e=0.95$. After 22.4 Myr, the primary becomes a HG star. Within the next 0.3 Myr its Roche lobe is overfilled and the binary's orbit becomes fully circularized. Until $t=22.5$ Myr, RLOF is ongoing and the binary parameters change drastically from: $M_1=10.2 M_{\odot}$, $R_1=17.5 R_{\odot}$, $M_2=9.0 M_{\odot}$, $a=44.78 R_{\odot}$ to $M_1=2.26 M_{\odot}$, $R_1=9.4 R_{\odot}$, $M_2=12.95 M_{\odot}$, $R_2=6.28 R_{\odot}$, $a=230.3 R_{\odot}$. From $t=22.4$ Myr to $t=26.65$ Myr primary passes through HG, CHeB, as well as MS, HG and giant branch as a naked He star, finally becoming an ONeMg WD. During these processes $a$ increased to $558.0 R_{\odot}$, $M_1$ dropped to 1.36 $M_{\odot}$. Between $t=30.1$ Myr and $t=32.5$ Myr, also the secondary passes through evolutionary channels as: HG, naked He star on MS, HG and finally giant branch naked He star. At the end of that period, RLOF starts on a thermal timescale. As a result, accretion-induced collapse occurs. Before the event, parameters of the binary were the following: $M_1=1.37 M_{\odot}$, $M_2=1.63 M_{\odot}$, $R_1=2.2\cdot 10^{-5} R_{\odot}$, $R_2=109.5 R_{\odot}$, $a=1.19 R_{\odot}$. As the neutron star was born, the mass and radius of the primary changed to: $M_1=1.26 M_{\odot}$, $R_1=9.74$ km. Other parameters did not change significantly.
\subsection{He WD}
\begin{figure}
\centering
\includegraphics[width=1.0\textwidth]{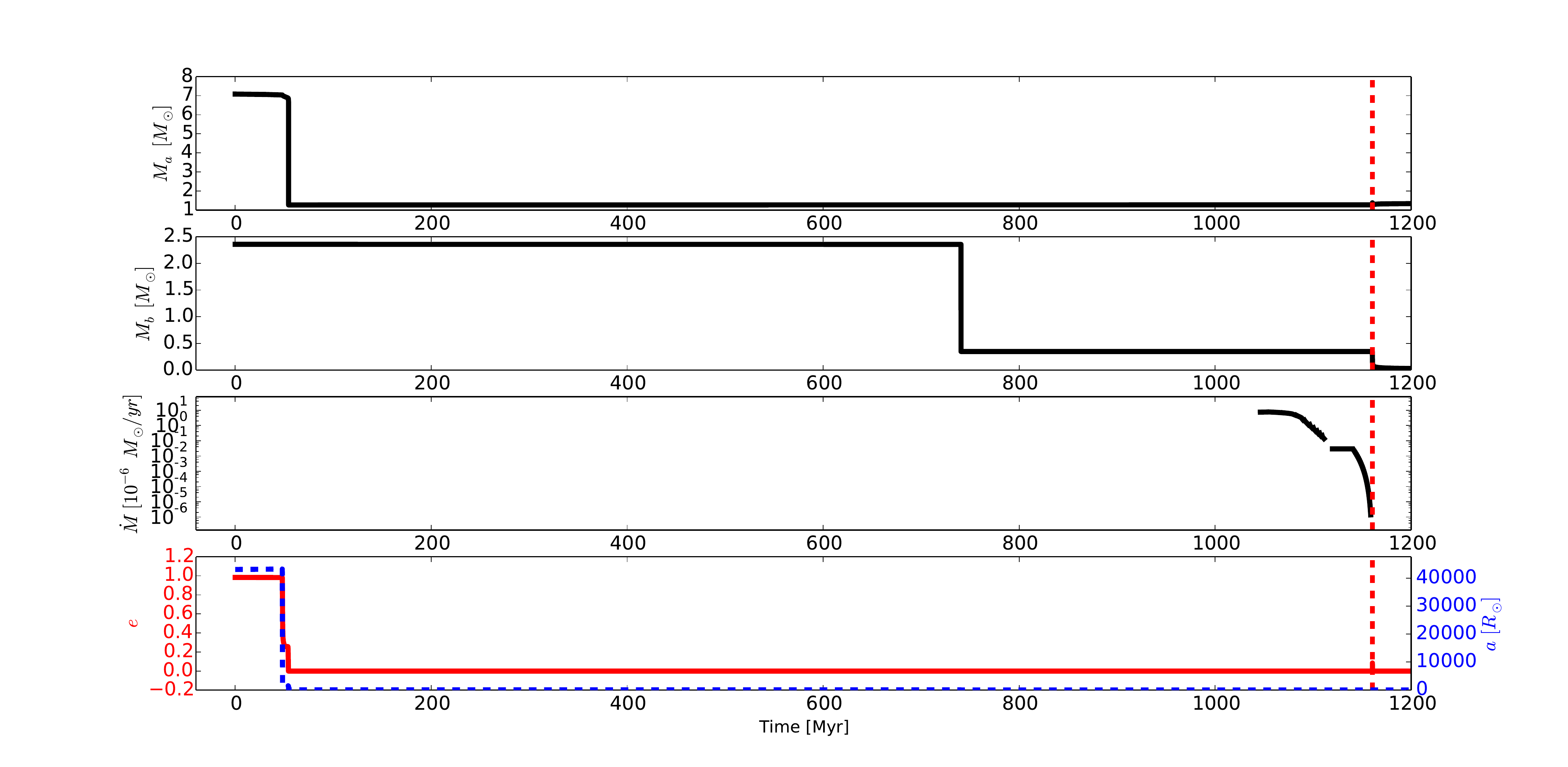}
\caption{Evolution of accretor mass $M_1(t)$, donor mass $M_2(t)$, mass transfer rate $\dot{M}(t)$, semi-major axis $a$ (blue dashed line) and eccentricity $e$ (red line) of an orbit for He WD donor. A red dashed vertical line represents a moment when AIC occurs.}
\end{figure}
Evolution starts with ZAMS stars of the following parameters: $M_1=7.1 M_{\odot}$, $R_1=7.6 R_{\odot}$, $M_2=2.36 M_{\odot}$, $R_2=1.8 R_{\odot}$. The semi-major axis is $a=4.31\cdot 10^4 R_{\odot}$ and eccentricity $e=0.98$. Between 47.8 Myr and 54.4 Myr, primary enters MS, HG, giant branch, CHeB, AGB and becomes a thermally pulsating AGB star. At the end of this period, also the orbit was circularized by tidal forces and finally primary became an ONeMg WD after a CE phase. From 728.9 Myr to 740.7 Myr, the secondary passes through MS, HG and giant branch. Finally, it becomes a He WD after ejecting the envelope from CHeB phase. At this point, binary parameters are the following: $M_1=1.27 M_{\odot}$, $M_2=0.35 M_{\odot}$, $R_1=2712$ km, $R_2=0.017 R_{\odot}$ and $a=1.18 R_{\odot}$. The mass transfer starts at $t=1.0$ Gyr. At $t=1.16$ Gyr, the binary passes through an AIC event and the primary becomes a neutron star. Also the mass transfer terminates at this point. Just prior to this, the mass of the primary was $M_1=1.38 M_{\odot}$ ($R_1=1.3\cdot 10^{-4} R_{\odot}$), mass of the secondary star was $M_2=0.16 M_{\odot}$ ($R_2=0.023 R_{\odot}$) and $a=0.1 R_{\odot}$. After the AIC, mass and radius of primary changed to $M_1=1.26 M_{\odot}$, $R_1=9.74$ km. 
\subsection{CO WD}
\begin{figure}
\includegraphics[width=1.0\textwidth]{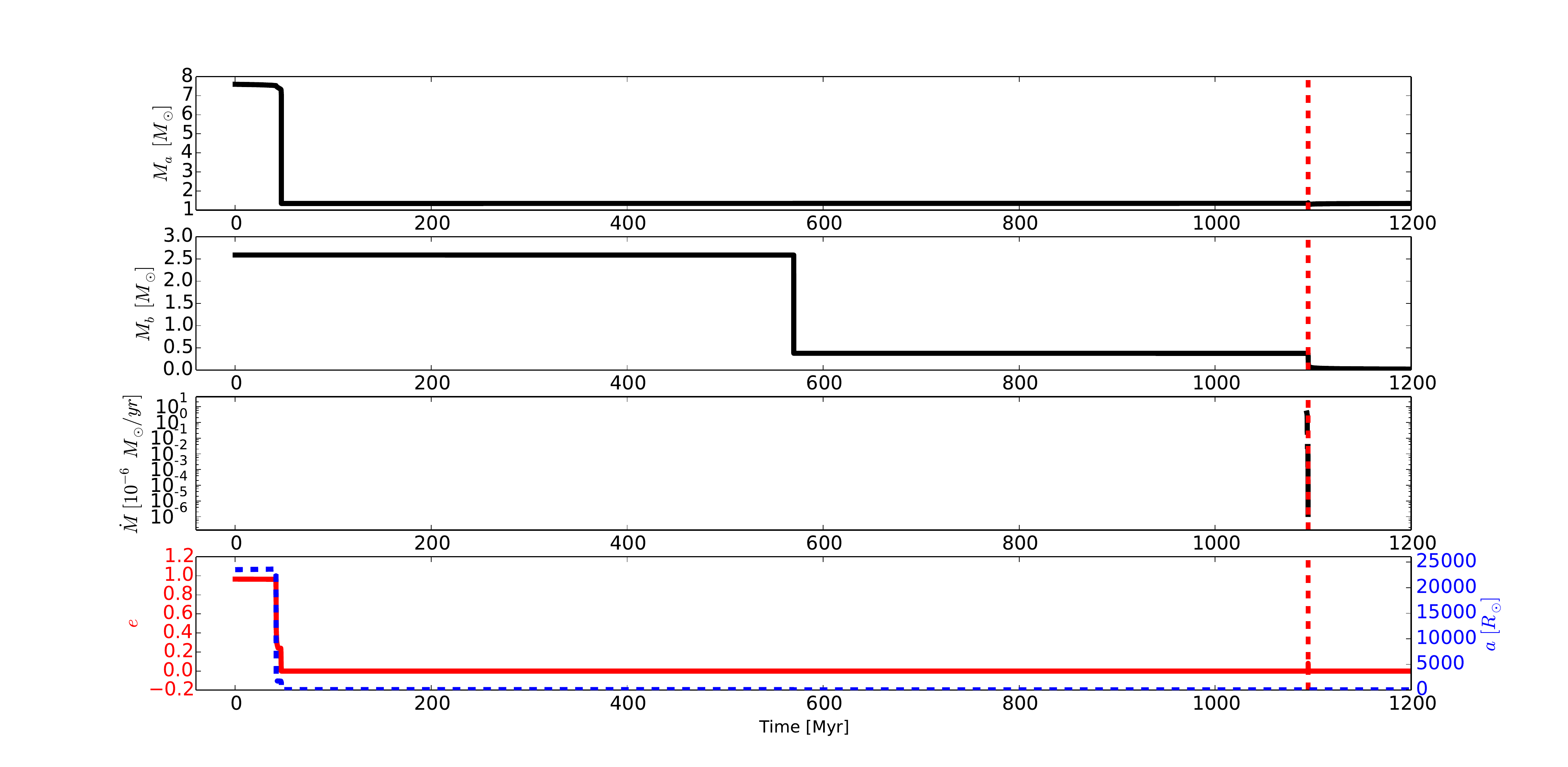}
\caption{Evolution of accretor mass $M_1(t)$, donor mass $M_2(t)$, mass transfer rate $\dot{M}(t)$, semi-major axis $a$ (blue dashed line) and eccentricity $e$ (red line) of an orbit for CO WD donor. A red dashed vertical line represents a moment when AIC occurs.}
\end{figure}
At ZAMS binary consists of stars of the following properties: $M_1=7.6 M_{\odot}$, $R_1=7.9 R_{\odot}$, $M_2=2.6 M_{\odot}$, $R_2=1.9 R_{\odot}$. Semi-major axis is $a=2.35\cdot 10^4 R_{\odot}$ and eccentricity $e=0.96$. Primary passes through MS, HG, giant branch and becomes an early AGB star between $t=41.4$ Myr and $ 46.7$ Myr. Shortly after that, the orbit was circularized by tidal forces. Then, the primary starts thermal pulsations and becomes an ONeMg WD through a CE phase. At $t=562.35$ Myr, the secondary enters the HG phase and in the next 4 Myr becomes a giant. After another 3 Myr, the star returns onto the MS as a naked He star. At $t=1017.71$ Myr, the star enters the HG and at 1056.35 Myr it becomes a CO WD. A thermal timescale mass transfer starts at $t=1094.70$ Myr. Accretion-induced collapse occurs at $t=1094.91$ Myr and mass transfer phase terminates at the same time. Before that, the primary mass was $M_1=1.38 M_{\odot}$ ($R_1=79.29$ km) and the secondary - $M_2=0.22 M_{\odot}$ ($R_2=0.02 R_{\odot}$). Orbital radius was $a=0.085 R_{\odot}$. Just after creation the neutron star has a mass $M_1=1.26 M_{\odot}$ and radius $R_1=9.74$ km. Other parameters have not changed significantly.
\begin{table*}
\centering
 \begin{minipage}{140mm}
  \caption{Statistics of mass transfer rates in AIC progenitor binaries for different galaxy metallicities: $100\%$, $10\%$ and $1\%$ of solar metallicity. \textbf{Donor type:} MS - main-sequence star, HG - Hertzsprung gap star, RG - early red giant, early AGB - early asymptotic giant branch star, pulsing AGB - thermally pulsing AGB star, naked He - naked He star, He WD - helium white dwarf, CO WD - carbon-oxygen white dwarf. \textbf{Symbols}: $\dot{M}_{avg}$ - mean mass transfer rate (averaged over time intervals when MT occurs), $\dot{M}_{max}$ - maximum mass transfer rate}
\begin{tabular}{ll|l|l|}
Metallicity $[Z_{\odot}]$    & Donor type         & $\dot{M}_{avg}$ [$M_{\odot}$/yr] & $\dot{M}_{max}$ [$M_{\odot}$/yr] \\ \hline
\multirow{11}{*}{$100\%$} &MS, $M\leq 0.7 M_{\odot}$&-&-\\
&MS, $M\geq 0.7 M_{\odot}$&7.4$\cdot 10^{-5}$&2.6$\cdot 10^{-4}$\\
&HG&1.2$\cdot 10^{-4}$&6.7$\cdot 10^{-4}$\\
&RG&2.3$\cdot 10^{-7}$&3.0$\cdot 10^{-6}$\\
&early AGB&1.8$\cdot 10^{-6}$&2.9$\cdot 10^{-6}$\\
&pulsing AGB&-&-\\
&MS, naked He&2.5$\cdot 10^{-8}$&3.7$\cdot 10^{-8}$\\
&HG, naked He&2.5$\cdot 10^{-7}$&4.2$\cdot 10^{-7}$\\
&GB, naked He&3.5$\cdot 10^{-3}$&2.1$\cdot 10^{-2}$\\
&He WD&9.5$\cdot 10^{-7}$&7.7$\cdot 10^{-6}$\\
&CO WD&4.7$\cdot 10^{-7}$&4.3$\cdot 10^{-6}$\\ \hline
\multirow{11}{*}{$10\%$}&MS, $M\leq 0.7 M_{\odot}$&-&-\\
&MS, $M\geq 0.7 M_{\odot}$&7.4$\cdot 10^{-5}$&2.6$\cdot 10^{-4}$\\
&HG&1.2$\cdot 10^{-4}$&6.7$\cdot 10^{-4}$\\
&RG&2.3$\cdot 10^{-7}$&3.0$\cdot 10^{-6}$\\
&early AGB&1.8$\cdot 10^{-6}$&2.9$\cdot 10^{-6}$\\
&pulsing AGB&-&-\\
&MS, naked He&2.5$\cdot 10^{-8}$&3.7$\cdot 10^{-8}$\\
&HG, naked He&2.5$\cdot 10^{-7}$&4.2$\cdot 10^{-7}$\\
&GB, naked He&3.5$\cdot 10^{-3}$&2.1$\cdot 10^{-2}$\\
&He WD&9.5$\cdot 10^{-7}$&7.7$\cdot 10^{-6}$\\
&CO WD&4.7$\cdot 10^{-7}$&4.3$\cdot 10^{-6}$\\ \hline
\multirow{11}{*}{$1\%$}&MS, $M\leq 0.7 M_{\odot}$&1.6$\cdot 10^{-9}$&2.6$\cdot 10^{-8}$\\
&MS, $M\geq 0.7 M_{\odot}$&7.4$\cdot 10^{-5}$&2.6$\cdot 10^{-4}$\\
&HG&1.2$\cdot 10^{-4}$&6.7$\cdot 10^{-4}$\\
&RG&2.3$\cdot 10^{-7}$&3.0$\cdot 10^{-6}$\\
&early AGB&1.8$\cdot 10^{-6}$&2.9$\cdot 10^{-6}$\\
&pulsing AGB&-&-\\
&MS, naked He&2.5$\cdot 10^{-8}$&3.7$\cdot 10^{-8}$\\
&HG, naked He&2.5$\cdot 10^{-7}$&4.2$\cdot 10^{-7}$\\
&GB, naked He&3.5$\cdot 10^{-3}$&2.1$\cdot 10^{-2}$\\
&He WD&9.5$\cdot 10^{-7}$&7.7$\cdot 10^{-6}$\\
&CO WD&4.7$\cdot 10^{-7}$&4.3$\cdot 10^{-6}$\\ \hline
\end{tabular}
\end{minipage}
\end{table*}

\pagebreak
\begin{figure}[ht]
\centering
\includegraphics[width=0.5\textwidth]{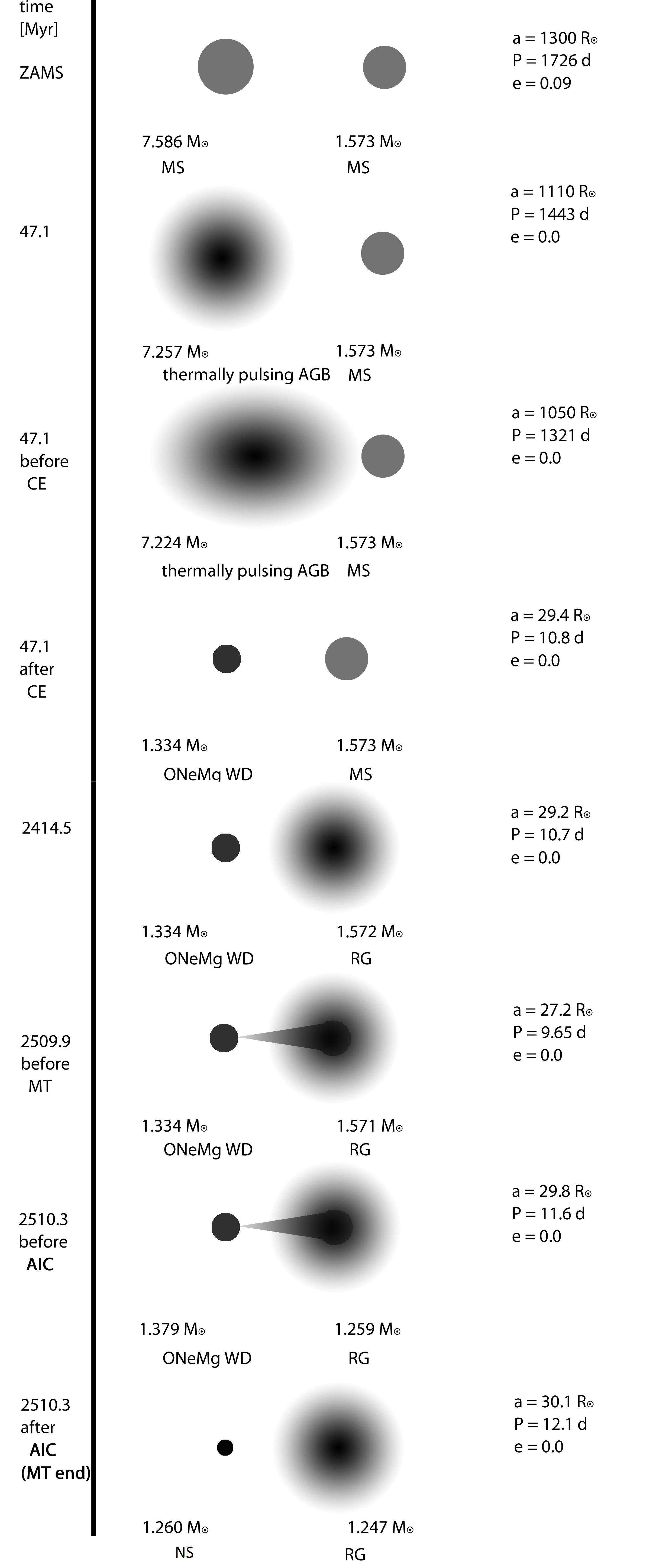}
\caption{Schematic picture showing evolution of AIC progenitor binary with an RG companion. Notation: MS - main sequence, RG - I giant branch star, AGB - asymptotic giant branch star, WD - white dwarf, NS - neutron star}
\end{figure}
\pagebreak
\begin{figure}[ht]
\centering
\includegraphics[width=0.5\textwidth]{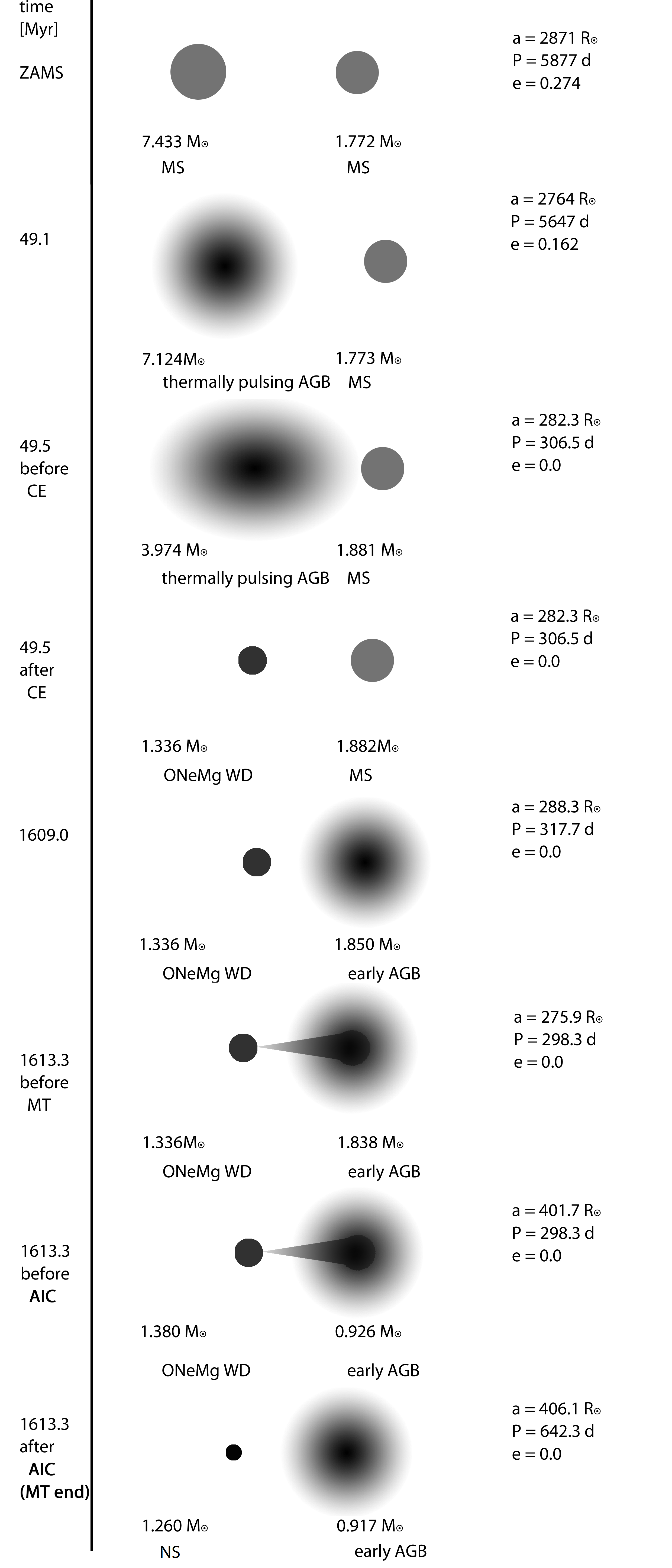}
\caption{Schematic picture showing evolution of AIC progenitor binary with early AGB stellar companion. Notation: MS - main sequence, RG - I giant branch star, AGB - asymptotic giant branch star, WD - white dwarf, NS - neutron star}
\end{figure}
\newpage
\begin{landscape} 
\begin{table}
\centering
 \begin{minipage}{140mm}
  \caption{Statistics of AIC events for different galaxy metallicities: $100\%$, $10\%$ and $1\%$ of solar metallicity. \textbf{Donor type:} MS - main-sequence star, HG - Hertzsprung gap star, RG - red giant, AGB - asymptotic giant branch star, He - He star, WD - white dwarf. \textbf{Symbols}: \% - percentage of this donor in a full sample of AIC progenitors, $M_{ZAMS}^{min,a}$ - lowest mass of accretor at ZAMS, $M_{ZAMS}^{max,a}$ - highest mass of accretor at ZAMS, $M_{ZAMS}^{min,b}$ - lowest mass of donor at ZAMS, $M_{ZAMS}^{max,b}$ - highest mass of donor at ZAMS, $t_{med}$ - median delay time in Myr, $t_{avg}$ - average delay time in Myr.}

\begin{tabular}{ll|l|l|l|l|l|l|l|l|l|}

Metallicity $[Z_{\odot}]$    & Donor type         &\%& $M_{ZAMS}^{min,a}$        & $M_{ZAMS}^{max,a}$ & $M_{ZAMS}^{min,b}$ & $M_{ZAMS}^{max,b}$ & $t_{med}$ [Myr] & $t_{avg}$ [Myr]\\ \hline
\multirow{6}{*}{$100\%$} &MS&2.64&7.43&10.90&1.30&1.96&729.88&744.58\\
&HG&1.40&7.49&10.68&1.20&2.46&902.22&1013.26\\ 
&RG&22.74&7.05&7.90&0.96&1.88&3227.22&4343.68\\ 
&AGB&30.94&7.12&7.84&0.93&7.63&83.41&191.31\\ 
&He&22.74&5.88&11.97&2.55&11.24&111.21&135.88\\
&WD&19.54&7.04&10.86&1.31&2.76&1579.37&1882.31\\ \hline
\multirow{6}{*}{$10\%$}&MS&0&-&-&-&-&-&-\\ 
&HG&12.69&6.05&10.18&1.06&3.12&600.84&728.54\\
&RG&14.56&6.00&7.14&0.81&2.35&2147.90&3060.95\\
&AGB&28.44&6.03&6.73&0.83&6.66&101.19&212.44\\ 
&He&22.59&6.01&11.98&2.12&11.02&134.8&169.14\\  
&WD&21.71&4.37&7.14&1.17&5.22&1206.01&1586.09\\ \hline
\multirow{6}{*}{$1\%$}&MS&1.20&5.74&6.34&0.12&1.29&6839.95&7061.05\\ 
&HG&28.24&5.59&6.35&0.83&3.15&447.83&644.41\\ 
&RG&25.66&5.59&6.30&0.80&2.98&2133.68&3169.21\\ 
&AGB&13.93&5.59&6.19&0.86&6.11&83.38&403.64\\ 
&He&19.28&5.58&9.80&2.30&8.34&230.05&246.63\\ 
&WD&11.69&4.08&6.23&1.32&5.06&1008.30&1046.59\\ \hline
\end{tabular}

\end{minipage}
\end{table}

\end{landscape}

\section{Delay Time Distribution}
In this part, I present the DTD for all progenitors of AIC events analyzed.
%As it was mentioned before, CE prescription used in these calculations is $\alpha\times\lambda=1$.
%Dependence on galaxy metallicity is investigated as well. 
%Evolution was allowed to proceed until 13.7 Gyr, which corresponds to the age of the Universe.
Statistical data are summarized in Table 2.2. In Figure 2.12 I present rates for elliptical galaxies, but since I assume an instantaneous burst of star formation at $t=0$ Gyr, the shape of the presented functions corresponds to the DTD.

\subsection{Solar metallicity}
The evolution leading to binary progenitor with MS secondary is described above. The DTD shows that these systems form within a rather narrow range of delay times with $t_{med}$=729.9 Myr. Due to this fact, ONeMg WD - MS progenitors are only 2.7\% of all AIC progenitors. ZAMS donors of three systems were found to have masses between 1.3 $M_{\odot}$ and 1.96 $M_{\odot}$ with primary masses between 7 $M_{\odot}$ and 10$M_{\odot}$. Mass transfer rates are quite low ($\dot{M}\sim 10^{-7} M_{\odot}yr^{-1}$) when compared to some SDS donors for SN~Ia progenitors, where H-rich matter is accreted onto a CO WD and accretion rates can be as high as $10^{-3} M_{\odot}yr^{-1}$. Binary partners of this composion are brought together predominantly by loss of orbital angular momentum from magnetic braking.\\ \\
Donors at the next phase of evolution, i.e. Hertzsprung gap, also make a small contribution to AIC progenitor scenarios - 1.4 \%. Primaries at ZAMS have 7-8 $M_{\odot}$ and secondaries 1.2-2.4 $M_{\odot}$. Generally, the collapse starts slightly later than in the case of MS secondaries, i.e. median time is $t_{med}=1.0$ Gyr and average - $t_{avg}=902.2$ Myr. Accumulation rates are of the order $\dot{M}\sim 10^{-7}-10^{-8} M_{\odot}yr^{-1}$.\\ \\

Red giant donors form a large collection of interesting systems, i.e. 22.74\%. Also, the behaviour of the DTD is much more complex than in prior systems. Most likely delay times are widely distributed between 1.8 Gyr and 10 Gyr. There are also systems, for which AIC occurs between 11.8-11.9 Gyr as well as even 12.7-12.9 Gyr after they were formed on ZAMS (median time - $t_{med}=3.2$ Gyr, average - $t_{avg}=4.3$ Gyr). Despite this fact, ZAMS masses of those stars are found in a very narrow range. Initial primary masses are found to be 7.05-7.90 $M_{\odot}$ and the secondary - from 0.96 $M_{\odot}$ to 1.88 $M_{\odot}$. The mass transfer rate is very high up to $\sim 10^{-3} M_{\odot}yr^{-1}$, on a thermal timescale, but hydrogen accumulation is fully efficient for a small range of accretion rates \citep{b35}. Most of the accreting binaries represent accumulation rates around $10^{-7} M_{\odot} yr^{-1}$.\\ \\
When the donor at the onset of AIC is an AGB star, the situation is even more complex. Even though those systems contribute to 30.94\% of all cases, being the highest percentage, the delay times are within a very narrow range. Most of systems start electron capture before first 100 Myr ($t_{med}=83.47$ Myr, $t_{avg}=191.31$ Myr). The oldest systems are found at $t=3 Gyr$. Primaries have a well defined mass at ZAMS: 7.12-7.84 $M_{\odot}$, but donors are found within a very wide range of ZAMS masses ($M_{ZAMS}=0.93-7.63 M_{\odot}$). Mass transfer occurs usually at $\dot{M}\sim 10^{-6}M_{\odot} yr^{-1}$, but roughly with 10\% efficiency.\\ \\
Helium stars are found as a likely donor of AIC progenitor binaries. They represent 22.74\% of systems. But again, collapse occurs when those stars are very young - median time $t_{med}=111.21$ Myr with a narrow and rapidly decaying DTD ($t_{med}=135.88$ Myr). In spite of this fact, ZAMS masses of both primaries ($M_{ZAMS}=5.88-11.97 M_{\odot}$) and secondaries ($M_{ZAMS}=2.55-11.24 M_{\odot}$) show a vast spectrum of possible progenitors. It is worth noting that within this scenario, I also enable for sub-Chandrasekhar mass SN~Ia explosions if the accretion rate and core mass of the primary ONeMg WD are within a certain range (more detailed description in \citet{b36} and study of He-accreting WDs in \citep{b39}). Roughly speaking, a sub-Chandrasekhar explosion is a possible outcome when the accretion rate is near $10^{-8} M_{\odot} yr^{-1}$ or lower.\\ \\
%Even though many systems are found within this accretion regime, none of them exploded as SN~Ia. What is more, all He-rich secondary progenitors of AIC pass through this region. -- I took this out, as it's puzzling to me. I don't know why - do you? there should be some systems in there. in any case, if you cannot explain why, then we need to leave this info out of the picture! if the referee asks about it then you'll have to investigate. maybe it has to do with the CE chosen prescription.\\ \\
I also find that 19.54 \% of AIC progenitors consist of WD donors, with either a He WD or a CO WD as a donor.
The primary forming on the ZAMS has a mass in the range 7.04 $M_{\odot}$ to 10.86 $M_{\odot}$ and secondary - 1.31 to 2.76 $M_{\odot}$. The majority of AIC events occur between 0.5-2.5 Gyr, but there are also several cases between 2.8-5.9 Gyr. As a result, the median time is $t_{med}=1579.37$ Myr and average - $t_{avg}=1882.31$ Myr. Evolutionary scenarios leading to AIC through this channel are described above.
\subsection{10\% of solar metallicity}
For a galaxy metallicity around $Z\sim 0.1 Z_{\odot}$, no progenitors with an MS donor have been found. Probably, this is due to the fact that in CE model used within this study, changing metallicity means also changing the $\lambda$ value.\\ \\
In case of Hertzsprung gap secondaries, which represent 12.69\% of all types of AIC progenitors, delay times are between 0.1 and 2.0 Gyr. As most of the AICs occur before 1 Gyr, the median time is $t_{med}=600.84$ Myr and average - $t_{avg}=728.54$ Myr. ZAMS masses of donors ($M_{ZAMS}=1.06-3.12 M_{\odot}$) and accretors ($M_{ZAMS}=6.05-10.18 M_{\odot}$) are very similar to those of solar metallicity galaxies.\\ \\
Red giant donors account for 14.56\% of all AIC progenitor binaries. The primary forming on the ZAMS has a mass between 6 $M_{\odot}$ and 7.14 $M_{\odot}$ and the secondary - from 0.81 to 2.35 $M_{\odot}$. Thus, both companions form in a rather narrow range of masses. Nevertheless, DTD shows a broad and slowly decaying distribution between 0.1 and 8.7 Gyr, with most of events around $t_{med}=2147.9$ Myr (average time: $t_{avg}=3060.95$ Myr).\\ \\
Again, AGB donors are found to be the most likely progenitor for AIC - 28.44\% of all events. The vast majority of these systems undergo AIC before 200 Myr ($t_{med}=101.19$ Myr, $t_{avg}=212.44$ Myr), but also substantial number of events occured between 1.5 and 3.5 Gyr. Primaries form within a narrow range of ZAMS masses, i.e. 6.03-6.73 $M_{\odot}$. In spite of this fact, secondaries at ZAMS can have 0.93 $M_{\odot}$ or be as massive as 6.66 $M_{\odot}$.\\ \\
Helium star donors are also very likely as they represent 22.59 \% of pre-AIC secondaries. ZAMS masses of these stars are between 2.12 and 11.02 $M_{\odot}$. Accretors for these progenitors form at ZAMS having masses from 6.01 up to even 12 $M_{\odot}$. Most AIC events occur after a very short delay time (median time - $t_{med}=134.8$ Myr and average time - $t_{avg}=169.14$ Myr). \\ \\
Degenerate secondaries also give a strong contribution to AIC scenarios (21.71\% of all progenitors). Even though they show a wide distribution of delay times, the median time is around 135 Myr and average - 169 Myr. Initial mass of donor is between 6.01 and 11.98 $M_{\odot}$. ZAMS masses of accretors are from 2.12 and 11.02 $M_{\odot}$.
\subsection{1\% of solar metallicity}
Main Sequence donors correspond to the lowest percentage of AIC progenitor binaries, i.e. 1.2\%. Their ZAMS mass is between 5.74 and 6.34 $M_{\odot}$. Accretors are also found within a narrow range, i.e. 0.12 - 1.29 solar masses. As a result, DTD shows a small amount of systems undergoing AIC between 4.1 and 5.2 Gyr, but also 9.0-9.1 Gyr. Thus, the median time is equal to 6839.95 Myr and average 7061.05 Myr.\\ \\
Hertzsprung gap donor channel represents 28.24 \% of all progenitors, being the most likely scenario for my lowest metallicity model. Primaries in these systems have ZAMS masses between 5.59 and 6.35 $M_{\odot}$. Initial masses of their companions are from 0.83 and 3.15 $M_{\odot}$. The DTD is a rather steep curve with delay times from 0.1 to 2.1 Gyr ($t_{med}=447.83$ Myr, $t_{avg}=644.41$ Myr). \\ \\
Red giant donors are also quite abundant among discussed binaries as they constitute 25.66 \% of all progenitors. What is more, ZAMS masses of both primary and secondary stars are the same as for the HG donor channel within this metallicity. Possible delay times are distributed among all feasible values from 100 Myr up to 13 Gyr. The DTD is exponentially decaying until 4.2 Gyr, where the function is roughly constant. Median time is $t_{med}=2133.68$ Myr and average time is $t_{avg}=3169.21$ Myr.\\ \\
In the case of the AGB donor channel, ZAMS masses of the primary are also within the same range as in the cases of RG and HG. Secondaries are found within a wider interval between 0.86 and 6.11 $M_{\odot}$. They represent 13.93\% of all progenitors. The DTD can be described by 2 rapidly decaying functions between 0.1-1 Gyr and 1.1-4 Gyr, with the former having a higher maximum value. As a result, median time for this channel is $t_{med}=2133.68$ Myr, but average - $t_{avg}=3169.21 Gyr$.\\ \\
Helium stars, as secondaries of pre-AIC binary systems, constitute 11.69\% of all available scenarios. ZAMS masses of primary stars are between 5.58 and 9.8 $M_{\odot}$. For secondaries, these values are found from 2.3 to 8.34 $M_{\odot}$. The DTD for these systems is a slowly decaying function for delay times up to 1.8 Gyr, but also with some systems for 2.3-2.5 Gyr. Thus, the median time $t_{med}=230.05$ Myr and average - $t_{avg}=246.63$ Myr.
\section{Rates}
\begin{table*}
\centering
 \begin{minipage}{140mm}
  \caption{Rates of AIC events for MW-mass spiral and elliptical galaxies of age 1 Gyr and 10 Gyr as a function of metallicity. 
\textbf{Donor type:} MS - main-sequence star, HG - Hertzsprung gap star, RG - red giant, AGB - asymptotic giant branch star, He - He star, WD - white dwarf}
\begin{tabular}{ll|l|l|l|l|}
                     &          & \multicolumn{2}{l|}{$t_{gal}$ = 1 Gyr} & \multicolumn{2}{l|}{$t_{gal}$ = 10 Gyr} \\ \cline{3-6} 
Donor type  & Metallicity [$Z_{\odot}$]  & elliptical        & spiral     & elliptical        & spiral      \\ \hline
\multirow{3}{*}{MS}&$100\%$&1.30$\cdot 10^{-5}$&2.66$\cdot 10^{-6}$&6.49$\cdot 10^{-6}$&3.05$\cdot 10^{-6}$\\
&$10\%$&1.62$\cdot 10^{-4}$&1.45$\cdot 10^{-5}$&6.49$\cdot 10^{-6}$&2.15$\cdot 10^{-5}$\\
&$1\%$&6.49$\cdot 10^{-6}$&6.49$\cdot 10^{-8}$&6.49$\cdot 10^{-6}$&1.62$\cdot 10^{-6}$\\ \hline
\multirow{3}{*}{HG}&$100\%$&2.60$\cdot 10^{-5}$&7.79$\cdot 10^{-7}$&6.49$\cdot 10^{-6}$&1.62$\cdot 10^{-6}$\\
&$10\%$&1.62$\cdot 10^{-4}$&1.45$\cdot 10^{-5}$&6.49$\cdot 10^{-6}$&2.15$\cdot 10^{-5}$\\
&$1\%$&1.49$\cdot 10^{-4}$&4.43$\cdot 10^{-5}$&6.49$\cdot 10^{-6}$&5.14$\cdot 10^{-5}$\\ \hline
\multirow{3}{*}{RG}&$100\%$&6.49$\cdot 10^{-6}$&0&1.30$\cdot 10^{-5}$&2.48$\cdot 10^{-5}$\\
&$10\%$&6.49$\cdot 10^{-6}$&6.49$\cdot 10^{-7}$&6.49$\cdot 10^{-6}$&2.39$\cdot 10^{-5}$\\
&$1\%$&4.54$\cdot 10^{-5}$&5.84$\cdot 10^{-7}$&6.49$\cdot 10^{-6}$&4.46$\cdot 10^{-5}$\\ \hline
\multirow{3}{*}{AGB}&$100\%$&1.30$\cdot 10^{-5}$&3.45$\cdot 10^{-5}$&6.49$\cdot 10^{-6}$&3.57$\cdot 10^{-5}$\\
&$10\%$&1.30$\cdot 10^{-5}$&4.63$\cdot 10^{-5}$&6.49$\cdot 10^{-6}$&4.83$\cdot 10^{-5}$\\
&$1\%$&1.30$\cdot 10^{-5}$&2.12$\cdot 10^{-5}$&6.49$\cdot 10^{-6}$&2.54$\cdot 10^{-5}$\\ \hline
\multirow{3}{*}{He}&$100\%$&6.49$\cdot 10^{-6}$&2.62$\cdot 10^{-5}$&6.49$\cdot 10^{-6}$&2.63$\cdot 10^{-5}$\\
&$10\%$&6.49$\cdot 10^{-6}$&3.83$\cdot 10^{-5}$&6.49$\cdot 10^{-6}$&3.83$\cdot 10^{-5}$\\
&$1\%$&1.30$\cdot 10^{-5}$&3.52$\cdot 10^{-5}$&6.49$\cdot 10^{-6}$&3.53$\cdot 10^{-5}$\\ \hline
\multirow{3}{*}{WD}&$100\%$&1.36$\cdot 10^{-4}$&1.56$\cdot 10^{-6}$&6.49$\cdot 10^{-6}$&2.26$\cdot 10^{-5}$\\
&$10\%$&2.98$\cdot 10^{-4}$&9.02$\cdot 10^{-6}$&6.49$\cdot 10^{-6}$&3.68$\cdot 10^{-5}$\\
&$1\%$&2.14$\cdot 10^{-4}$&8.37$\cdot 10^{-6}$&6.49$\cdot 10^{-6}$&2.14$\cdot 10^{-5}$\\ \hline
\end{tabular}
\end{minipage}
\end{table*}

\begin{figure}
\centering
\includegraphics[width=1.0\textwidth]{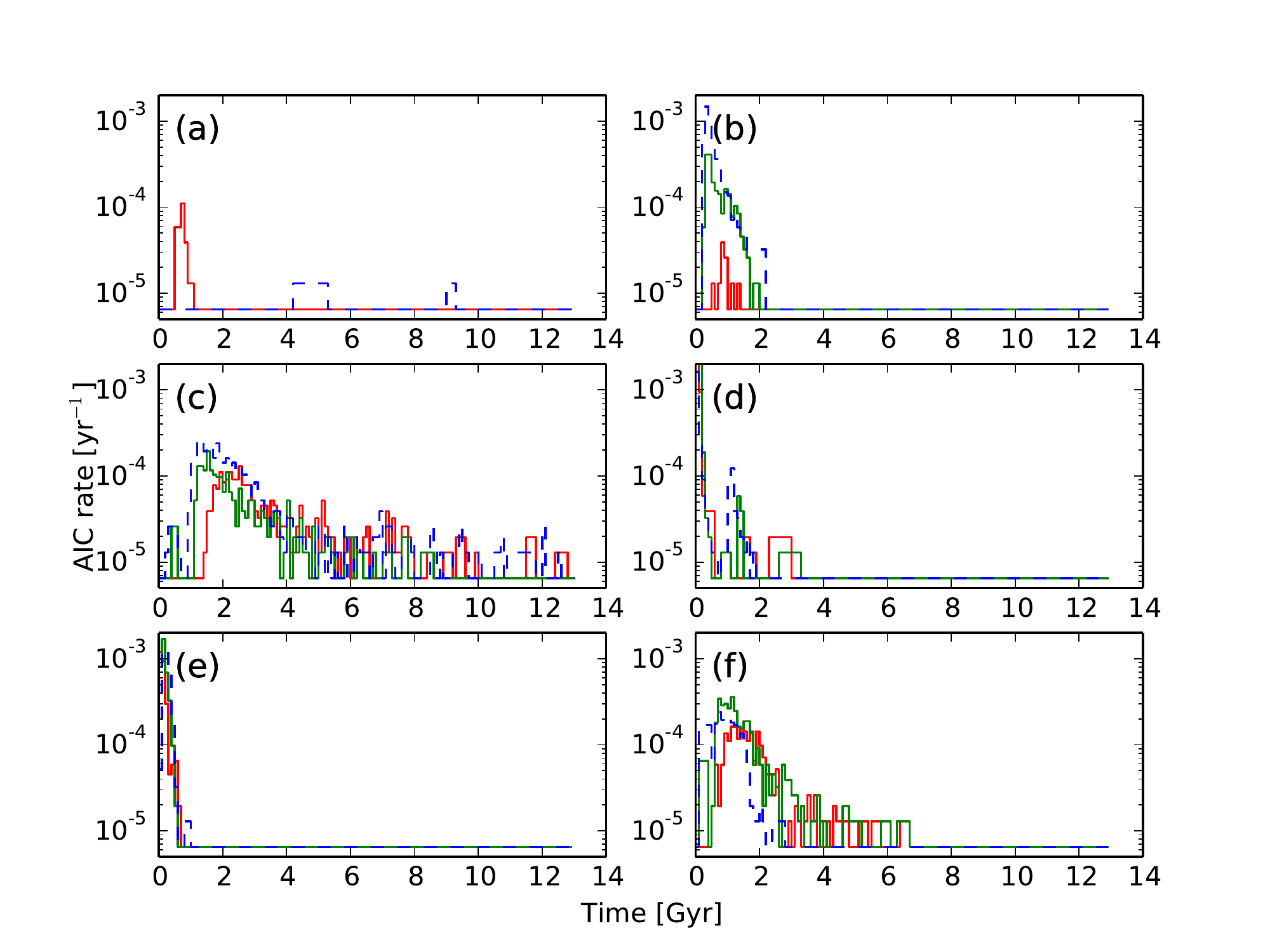}
\caption{Rates of AIC events for ellipticals. \textbf{Donor type:} (a) MS star, (b) HG star, (c) RG star, (d) AGB star, (e) He star, (f) WD star. Red line represents results for $Z=Z_{\odot}$, green - for $Z=0.1Z_{\odot}$ and blue dashed line for $Z=0.01 Z_{\odot}$}
\end{figure}
\begin{figure}
\centering
\includegraphics[width=1.0\textwidth]{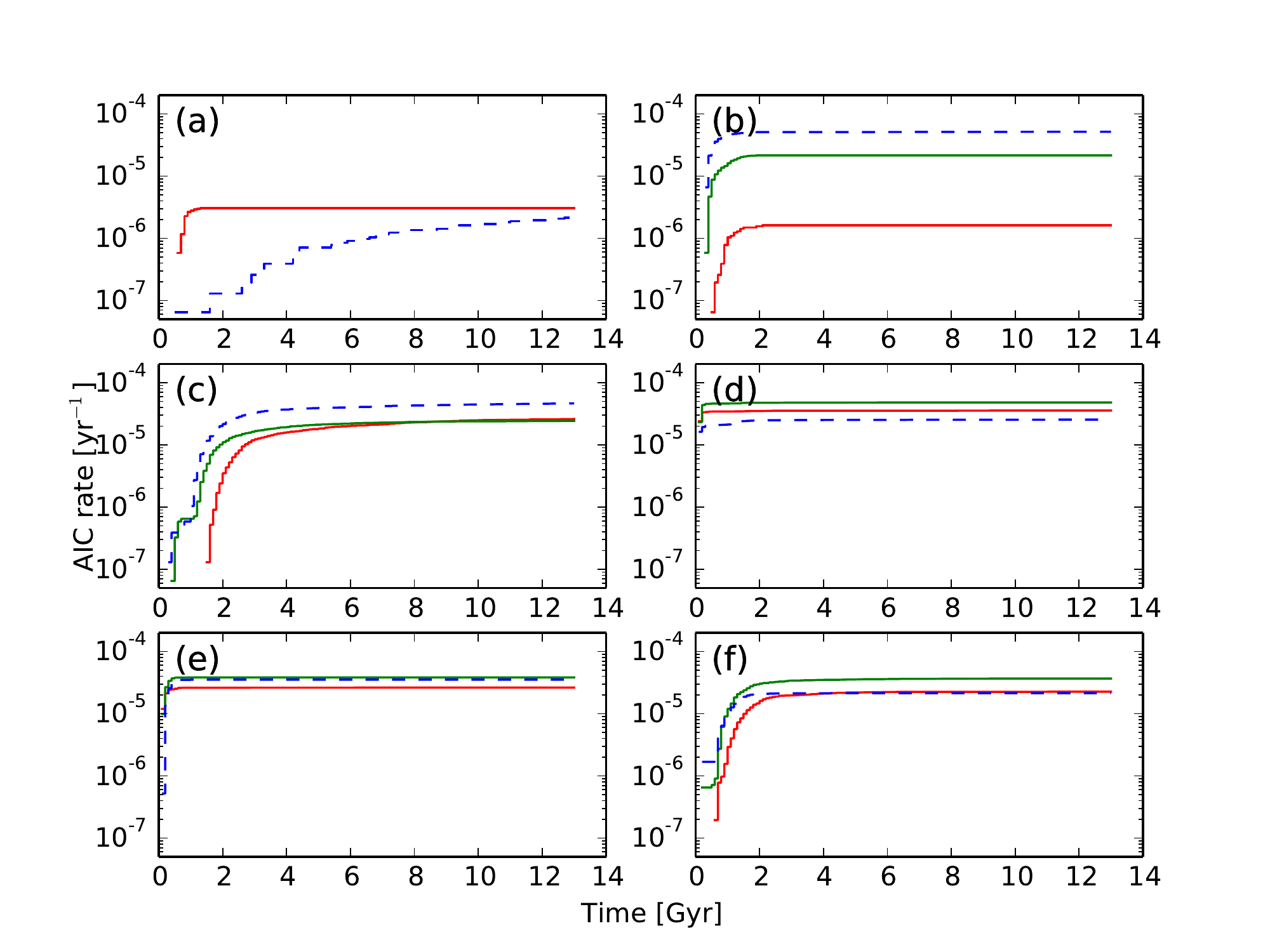}
\caption{Rates of AIC events for spirals. \textbf{Donor type:} (a) MS star, (b) HG star, (c) RG star, (d) AGB star, (e) He star, (f) WD star. Red line represents results for $Z=Z_{\odot}$, green - for $Z=0.1Z_{\odot}$ and blue dashed line for $Z=0.01 Z_{\odot}$}
\end{figure}
I assume a binary fraction of 50\% for analyzed synthetic galaxies. As a result, 0.032\% of all stellar systems undergo AIC. This observation can be applied to both types of galaxies, as they only differ in SFR. Rates of AIC events - $\mathcal{R}(t_{gal})$ - are presented in Figure 2.12 for elliptical galaxy (instantaneous starburst at $t=0$) and in Figure 2.13 for spiral (constant SFR $=6M_{\odot}yr^{-1}$ until $t=10$ Gyr). I show the rates in $yr^{-1}$ for a MW-mass galaxy, i.e. per 6 $\odot 10^{10} M_{\odot}$ bound in stars at present. Results for different donors in progenitor binaries, as well as metallicities, i.e. $Z=Z_{\odot},~0.1Z_{\odot},~0.01Z_{\odot}$ are presented separately in Table 2.3.
\subsection{Main Sequence donors}
Rates for elliptical galaxies are found within a very narrow range around 0.9-1.1 Gyr. For solar metallicity, the observed rate for a 1 Gyr old galaxy is equal to $1.30\cdot 10^{-5}~yr^{-1}$, but if the age is around 10 Gyr, the rate is very small $\sim 10^{-6}$. If the metallicity is 10 times smaller the rate is around $10^{-6} yr^{-1}$ for either galaxy age. If the smallest investigated metallicity, i.e. $Z=0.1Z_{\odot}$, is assumed, the rates are equal to $10^{-5} yr^{-1}$, for ages between 4.2 and 5.6 Gyr, but also at 9.1 Gyr. For any other times the rate is around $6.49\cdot 10^{-6}yr^{-1}$. \\ \\
MS donors for spiral galaxies are also very rare; for every metallicity tested there are very few cases found. Rates are found between $10^{-8} - 10^{-6}~yr^{-1}$. Functions presented on Figure 2.13 d) show different behaviour for different metallicity. If $Z=Z_{\odot}$ or $Z=0.1 Z_{\odot}$, the function is rapidly growing from 1.1 to 1.4 Gyr with rates between $4.5 \cdot 10^{-6}yr^{-1}$ and $8\cdot 10^{-5} yr^{-1}$ and then remains constant. For $Z=0.1Z_{\odot}$ the rate is a slowly increasing function of galaxy age until 5 Gyr. Values are in the range from $6.49 \cdot 10^{-8} yr^{-1}$ to $1.62 \cdot 10^{-6}$. Rates for older galaxies are roughly constant at the level of $\sim 10^{-6} yr^{-1}$.
\subsection{Hertzsprung gap donors}
As it was mentioned in a prior section, this type of donor is quite abundant in elliptical galaxies of lower metallicity than $Z_{\odot}$. Additionally, the rates drop exponentially when presented as a function of galaxy age. Data also show a very distinct dependence on metallicity, but with relevant values for age up to 2.1 Gyr. If $Z=Z_{\odot}$, rates are around $4\cdot 10^{-5}$ and $8\cdot10^{-7}$ for galaxies roughly at the age between 0.4 and 1.6 Gyr. When $Z=0.1Z_{\odot}$, the highest rates are between $4\cdot 10^{-4}$ for age $t_{gal}=0.3$ Gyr and $1.5 \cdot 10^{-6} yr^{-1}$ for $t_{gal}=2.1$ Gyr. Then again, the rate is roughly constant. Exponential decay of $\mathcal{R}(t_{gal})$ is between $10^{-3} yr^{-1}$  for 0.3 Gyr and $3\cdot 10^{-5} yr^{-1}$. For older galaxies the AIC rate for this type of progenitor binary is around $\sim 6\cdot 10^{-6} yr^{-1}$. \\ \\
Dependence on metallcity is also evident in spiral galaxies. The shape of $\mathcal{R}(t_{gal})$ is exactly the same for different metallicities, but the range of values is totally different. Rates rapidly grow until $t_{gal}=1 Gyr$ and then the growth becomes slower until $t_{gal}=1.9$ Gyr, and then remain constant. In the case of a solar metallicity, rates range from $ 6\cdot 10^{-8} yr^{-1}$ up to $3.05 \cdot 10^{-6} yr^{-1}$. For metallicity of $10\% Z_{\odot}$, the rates are between $5.5 \cdot 10^{-7} yr^{-1}$ and $2.15\cdot 10^{-5} yr^{-1}$. If $Z=0.01 Z_{\odot}$, the examined function reaches values from $6.5 \cdot 10^{-6} yr^{-1}$ up to $5.14 \cdot 10^{-5} yr^{-1}$.
\subsection{Red giant donors}
Roughly speaking, different metallicities of galaxies for this type of donor change neither shape of $\mathcal{R}(t_{gal})$ nor rates of AIC events from this progenitor channel. When averaged over Monte Carlo noise, this is particularly true for $Z_{\odot},~0.01 Z_{\odot}$. AIC rates grow to $2.5 \cdot 10^{-4} yr^{-1}$ for the youngest galaxies up to 2 Gyr. Then, until $t_{gal}=5$ Gyr, rate values drop exponentially to $2.5-4.5 \cdot 10^{-5} yr^{-1}$. After this time the rate generally remains constant for the remaining epochs.\\ \\
In spiral galaxies, metallicity dependence is more complex than in prior channels. When $Z=Z_{\odot}$, nonzero rates appear at $t_{gal}=1.7$ Gyr and the value is as low as $1.3\cdot 10^{-7} yr^{-1}$. Rapid growth of rate-age dependence continues until $\mathcal{R}(t_{gal}=2.7$ Gyr $)\approx 10^{-5} yr^{-1}$. When ages are lower than 7.4 Gyr, rates slightly grow to 5 events occuring within 20,000 years and remain constant for older spirals. In the case of either 10\% and 1\% of solar metallicity, rates rapidly rise until $t_{gal}=2.8$ Gyr. In between there is an interval (0.8-1.2 Gyr) for which values are constant ($\mathcal{R}(t_{gal})=6\cdot 10^{-7} yr^{-1}$). In older galaxies, of $Z=0.1 Z_{\odot}$, rates are constant and equal to $2.39\cdot 10^{-5} yr^{-1}$. In the case of $Z=0.01 Z_{\odot}$, AIC from this channel is at least twice as likely. 
\subsection{AGB donors}
Even though for ellipticals this type of donor is the most likely, $\mathcal{R}(t_{gal})$ is a very steep function. With its values ranging from $5\cdot 10^{-3} yr^{-1}$ to $6.49\cdot 10^{-6} yr^{-1}$ the galactic ages are within 0.1-3.1 Gyr interval. More importantly, this description applies to each metallicity with rates slightly lower for increasing metallicity.\\ \\
The function that governs $\mathcal{R}(t_{gal})$ for spiral galaxies is also equivalent in terms of metallicity. A rapid growth can be observed within 0.2 Gyr from 0 to $\sim 10^{-5} yr^{-1}$. Then, values remain constant up to the age of the Universe. Differences between the highest rates are very low, i.e.: $4\cdot 10^{-5} yr^{-1}$ for $Z=Z_{\odot}$, $5\cdot 10^{-5} yr^{-1}$ for $Z=0.1Z_{\odot}$ and $2\cdot 10^{-5} yr^{-1}$ for $Z=0.01 Z_{\odot}$.
\subsection{He star donors}
Generally speaking, properties of this rate-galaxy age dependence are very similar to those of AGB donor channel. When looking at ZAMS masses of both AGB and He donors this is not a surprising result. AIC events from these binaries occur very frequently when the galaxy is just born, i.e. when the age is less than 0.2 Myr. Rates are of the order $1.7 \cdot 10^{-3} yr^{-1}$ for $0.1Z_{\odot}$ or $0.01 Z_{\odot}$. For solar metallicity, the rate at this age is slightly lower - $7\cdot 10^{-4} yr^{-1}$. For galaxy ages starting from 0.2 Gyr up to 0.9 Gyr the value drops exponentially to $6.5 \cdot 10^{-6} yr^{-1}$ for each metallicity. In case of $Z=0.01Z_{\odot}$ an additional peak was observed between 1.0-1.1 Myr when the rate reached $1.3\cdot 10^{-5} yr^{-1}$, but in older galaxies the rate does not exceed $6.5 \cdot 10^{-5} yr^{-1}$ for each metallicity.\\ \\
In the case of spirals, rates grow from 0 to $\sim 10^{-5} yr^{-1}$, when galaxies are younger than 0.2 Gyr. Then, values remain constant up to the age of the Universe. For solar metallicity this value is equal to $2.63\cdot 10^{-5}$, for 10\% - $3.83 \cdot 10^{-5}$ and for 1\% - $2.84 \cdot 10^{-5} yr^{-1}$. 
\subsection{WD donors}
This channel, when occurring in different ellipticals, exhibits very interesting characteristics. If $Z=Z_{\odot}$ the function of the rate versus age grows rapidly from $6.5 \cdot 10^{-6} yr^{-1}$ for $t_{gal}=0.4$ Gyr, up to $1.36\cdot 10^{-4} yr^{-1}$. Then it remains constant until 2 Gyr and starts to decline until 2.3 Gyr when the rate is equal to $10^{-5} yr^{-1}$ Then it remains constant until 5.9 Gyr. In galaxies of 10\% metallicity the situation is slightly different. Within a very short interval (0.1-0.4 Gyr) the rate is equal to $7 \cdot 10^{-5} yr^{-1}$. After that the function starts to rise rapidly to $3 \cdot 10^{-4} yr^{-1}$ for 1.8 Gyr. Subsequently, age dependence drops to $10^{-5} yr^{-1}$ and stays constant until 6.5 Gyr. For the lowest metallicity, the behaviour of this function can be summarized as follows: the rate remains constant until 0.8 Gyr having a value of $2\cdot 10^{-4} yr^{-1}$. Thereafter, the function starts to drop until it reaches $10^{-5} yr^{-1}$ at $t_{gal}=2.1$ Gyr.\\ \\
Pre-AIC double-degenerate systems start to occur in spiral galaxies of solar metallicity at 0.6 Gyr. Then $\mathcal{R}(t_{gal})$ grows to $9\cdot 10^{-5} yr^{-1}$ and remains constant from 2.9 Gyr until the age of Universe. Galaxy of 10\% metallicity has the same functional behaviour as the solar one, but the growth is ongoing from $6.5 \cdot 10^{-7} yr^{-1}$ to $2.5 \cdot 10^{-5} yr^{-1}$. Lowest metallicities galaxies show a very interesting case when there are 2 intervals when the function is similar to one of the metallicities, i.e. until 1.6 Gyr it follows $Z=0.1 Z_\odot$ and from 1.6 Gyr up to 13.7 Gyr it has a value identical to that of solar metallicity, $2\cdot 10^{-5} yr^{-1}$.
\chapter{Comparison with other studies}
During the last 20 years, there was in fact only one population synthesis study of AIC, that distinguishes between different type of donors in progenitor binaries. Table 2.1 in \citet{b39} presents birth rates for low-mass semidetached binaries. The He star donor channel would be a source of 13 events per 10,000 years.
%Such a value cannot be reached, due to observationally derived constraints for AIC rates that should be lower than $10^{-4} yr^{-1}$ for MW-type galaxy.
Such a value would be too high to reconcile with observationally-derived constraints for AIC rates: lower than $10^{-4} yr^{-1}$ for MW-type galaxies. 
This rate estimate is due to the presence of neutron-rich isotopes in the Solar System \citep{b55}. Another constraint emerging from nucleosynthesis study is presented in \citet{b64}. Authors claim that only $\lesssim$ 0.1\% of total Galactic neutron star population can be formed via AIC. Despite this discrepancy, rates provided in \citet{b39} agree with results presented in this work.\\ 
Generally speaking, other works concentrate on constraining and testing different CE prescriptions. One highly cited rate of AIC events is $\sim 8\cdot10^{-7} - 8\cdot 10^{-5} yr^{-1}$ for a synthetic Galaxy \citep{b20}. \citet{b26} imply that I can extrapolate the AIC rate from the observed SN~Ia rate $\sim 3\cdot 10^{-3} yr^{-1}$ \citep{b57,b58,b59}. The resulting value is $\sim 1.5 \cdot 10^{-4} yr^{-1}$ for my Galaxy. \citet{b64} provides a wider range of rates between $\sim 10^{-8} - 10^{-5} yr^{-1}$. \citet{b01} show I can also roughly estimate 13 AIC events per year, that would have been detected within 300 Mpc. \citet{b76} shows that with only a few events per year, gravitational wave detection from rapidly rotating proto-NSs would be difficult to achieve. \\ \\ 
Accretion-induced collapse from single degenerate channel is a natural by-product of the evolution starting at $M_{ZAMS,a}\sim 6-8M_{\odot}$ or up to $\sim 10 M_{\odot}$ \citep{b60}. Red giant donor in progenitor binary is thought to be the most effective scenario \citep{b01}. Authors also provide $M_{ZAMS,b}=0.9 M_{\odot}$ which agrees with my results. \\ \\
Another important parameter is mass transfer rate that allows to accrete enough matter for AIC to occur. \citet{b101} show that $\dot{M}\sim 10^{-8} M_{\odot} yr^{-1}$ is needed to accumulate He-rich matter onto ONeMg WD and induce AIC. H-rich matter accretion prescription used in this work is consistent with observations and it is indicated from H-R diagram of supersoft X-ray sources \citep{b35}.
\chapter{Discussion}
Population synthesis results presented in this work were derived from the evolution of single and binary stars using \texttt{StarTrack}. A more detailed study of each evolutionary channel in terms of donor type at the onset of AIC event has been presented. I have also analysed the resulting delay times and occurence rate of each progenitor type as well as rates for both elliptical and spiral MW-mass galaxies. Dependence on metallicity was also treated as a relevant parameter for future observations, and rates are summarized in Tables 2.2 and 2.3. CE parametrization, used in this study, is the commonly-adopted standard for the study of low- and intermediate-mass ZAMS binaries, i.e. $\alpha_{CE}\times\lambda=1.0$.\\ \\
My study was based on the assumption that ONeMg WDs steadily accumulate mass until some critical mass $M_{ecs}=1.38 M_{\odot}$ is reached and then an AIC occurs \citep{b18}.
%What is more, \citet{b65} find that there are cases of binary systems with ONeMg WD primary that explode as SNIa. Fortunately, it does not change results presented in this paper, because all of those SNIa progenitor donors explode before reaching $1.38 M_{\odot}$. Nevertheless, -- got rid of this, I indicated you should cite Marquardt et al. elsewhere. the way this is worded here is misleading - better to leave it out, or expand. 
There is also the possibility of a more thorough test of competition between explosion due to electron degeneracy and electron capture, using $\rho_{crit}$ for the nuclear reaction initiation. If one applies the values of critical densities from the literature, it turns out that the corresponding critical mass is approximately the same for each process. \citet{b18} also provide a very important estimate of AIC rates in globular clusters. As a result, this phenomenon can strongly account for the amount of MSPs. It is particularly important because I still lack thorough explanation for observed number of these pulsars. But if one scales the given values to MW mass just to compare, these are consistent with my rate estimates within an order of magnitude. This is a very promising result. \citet{b26} find that production of the MSPs spin is simple, when the orbital momentum conservation is assumed. \citet{b62} present a table of birth rates for post-AIC NS candidates for MSPs (Table 2.1 in that thesis). It turns out that electron capture processes can also account for the existence of MSPs in the Galactic disk. That is because the observed systems are presented as candidates for post-AIC NSs. The authors, though, notice that it is hard to observationally distinguish between post-AIC pulsars and those formed via standard SNIb/c. Fortunately, there is a possible solution. One can differentiate between He WD donors and RG donors, if the orbital periods of only post-AIC systems are 10-60 d and more than 500 d, respectively. Also velocities of MSPs are $\leq 30$ km/s which agrees well for GC pulsars from the AIC hypothesis. ZAMS masses of typical binary components as well as delay times agree with my estimates.\\ \\
I expect both long and short delay times for AICs. To be more accurate, the vast majority of events occur at delay times that are $t\sim 1-2$ Gyr. Unfortunately, there was no direct detection of any AIC event so far. For $Z=0.02 Z_{\odot}$ and $0.002 Z_{\odot}$, the AGB donor is a most likely progenitor. In the case of $Z=0.0002 Z_{\odot}$ the most likely scenario involves Red Giants which is also very abundant for other metallicities. 
Future investigations of AIC progenitors should also take into account the recently-proposed scenario for rotation-delayed AIC \citep{b63}. A typical evolutionary channel involves a ZAMS binary of masses $M_{ZAMS,a}=7.0 M_{\odot},~M_{ZAMS,b}=2.0 M_{\odot}$. After a RLOF and a CE phase, a binary consisting of a 1.2$M_{\odot}$ ONeMg WD and a 2.0$M_{\odot}$ MS donor is found. Then, rotation of such a system, as the system becomes a CV, prevents the WD from collapse and it does not pass through electron capture processes when it approaches the Chandrasekhar mass limit. A super-$M_{Ch}$ WD forms ($M_{ONeMg, superCh}=1.43 M_{\odot}$). The donor becomes a He WD of $M_b=0.27 M_{\odot}$. A Rotation-Delayed Accretion Induced Collapse would occur and could explain MSP+He WD systems. Thus, such a scenario is worth studying through population synthesis in the near future.

\end{document}